\documentstyle[aps]{revtex}
\begin{document}
\draft

\title{Non-equilibrium quantum noise in chiral Luttinger liquids}

\author{C. de C. Chamon}
\address{Department of Physics, Massachusetts
Institute of Technology, Cambridge, MA 02139}
\author{D. E. Freed}
\address{The Mary Ingraham Bunting Institute, Radcliffe
Research and Study Center, Harvard University, Cambridge, MA 02138}
\author{X. G. Wen}
\address{Department of Physics, Massachusetts
Institute of Technology, Cambridge, MA 02139}
\maketitle
\begin{abstract}
We study non-equilibrium noise in Chiral Luttinger Liquids using
the Landauer-Buttiker Scattering approach, obtaining the
current/voltage noise spectrum for a four-terminal measurement
scheme. Experimental consequences of the tunneling of charges are
present in the four-terminal measurement of both the low-frequency
shot noise ($\omega$ near 0), and the high-frequency Josephson
noise ($\omega$ near $\omega_J=e^*V/\hbar$).  Within perturbation
theory, an algebraic singularity is present (to all orders) at the
Josephson frequency $\omega_J=e^*V/\hbar$, whose position depends
on the charge $e^*$ of the tunneling particles, either electrons or
fractionally charged quasiparticles. We show in a non-perturbative
calculation for an exactly solvable point that the singularity at
the quasiparticle frequency exists only in the limit of vanishing
coupling, whereas the singularity at the electron frequency is
present for all coupling strengths. The vanishing coupling limit
corresponds to perfectly quantized Hall conductance in the case of
quasiparticle tunneling between edge states in the fractional
quantum Hall regime, and thus tunneling destroys the singularity at
the quasiparticle frequency concomitantly with the quantized
current.
\end{abstract}

\pacs{PACS: 72.10.Bg, 73.20.Dx, 73.40.Gk, 73.50.Fq, 73.50.Td}

\section{Introduction}

Recently it was realized that a strongly correlated 1D system,
namely a Chiral Luttinger liquid ($\chi$LL), exists in the edges of
fractional quantum Hall (FQH) liquids \cite{XGWcll}. Because of
their chiral nature, {\it i.e.}, the excitations in a given branch
move only in one direction, spatially separated branches cannot
interfere and cause localization of states. In contrast, non-chiral
Luttinger liquids are extremely sensitive to the presence of even
the smallest amounts of impurity in the sample, for in 1D all
states are localized, and long enough wires will behave as
insulators. The characteristic property of (chiral) Luttinger
liquids is that the tunneling conductance between the edge states
has a power law dependence on the temperature $\sigma\propto
T^{2(g-1)}$, where $g$ depends on the filling factor $\nu$ of the
FQH state, taking the values $g=\nu$ or $g=\nu^{-1}$ depending on
the tunneling geometry
\cite{XGW2,Kane&Fisher1,F&N,CCC&XGW}. By experimentally studying the
tunneling between edge states in the FQH regime using a point
contact geometry, Milliken, Umbach, and Webb \cite{Webb} found
this type of power law dependence of the tunneling conductance
on the temperature.
Their finding is consistent with the theoretical prediction $\sigma
\propto T^4$ for the $\nu=1/3$ FQH state \cite{XGW2,Moon}.

The experimental confirmation of the Luttinger liquid behavior in
tunneling between edge states has boosted theoretical interest in
further studies of properties of the conductance
\cite{Guinea1,Mak,Egger}. An exact solution for the conductance has
been obtained using the thermodynamic Bethe Ansatz, and an exact
duality between the $g$ and $1/g$ cases has been shown
\cite{Fendley1,Fendley2} in the context of the tunneling current,
as suggested in Ref. \cite{Duality}.
The rich behavior of tunneling in chiral Luttinger liquids extends
well beyond transport measurements alone
\cite{Kane&Fisher2,CFW,Fendley3}. One should expect, based on
experience in non-interacting systems, that the noise spectrum
contains information not attainable, in the most general case, from
just transport measurements. In general, the shape of the noise
spectrum is determined by the dynamical properties of the system,
which in turn contain information about the excited states. Even for
non-interacting electronic systems, non-trivial structures appear in
the noise \cite{Landauer,Buttiker,Lesovik}, the simplest example being
the suppression of classical shot noise due to quantum statistics. In
chiral Luttinger liquids, the tunneling particles sometimes carry
fractional charge and fractional statistics, and thus such strongly
correlated 1D systems also provide the natural experimental
realization for the study of features that arise in the noise spectrum
for generalized quantum statistics.

In reference \cite{CFW}, the noise spectrum of the tunneling current
between edge states directly at the point contact was calculated
perturbatively.  To low orders in the tunneling amplitude, we found
that there was a singularity at $\omega = 0$; for small $\omega$ the
noise spectrum has the form $S_{SN}+S_{\rm sing}(\omega)$, where
$S_{SN}$ is the zero-frequency shot noise and $S_{\rm
sing}(\omega)=c|\omega|$. The slope $c$ of the $|\omega|$ singularity
has a strong non-linear dependence on the applied voltage $V$
($c\propto (2g-1)^2V^{4(g-1)}$), which is another signature of
Luttinger liquid behavior (to be contrasted with the case of
non-interacting electrons, $g=1$, where the slope is independent of
$V$). The exponent $g$ characterizes the Luttinger liquid
behavior. This low-frequency part of the spectrum is the one more
easily accessible experimentally. Secondly, there is another
singularity at $\omega = \omega_J$ where $\omega_J = e^*V/\hbar$ is
the Josephson frequency of the electron ($e^*=e$) or quasiparticle
($e^*=\nu e$) that tunnels through the point contact.  The shape of
this singularity depends on $g$ and goes as $|\omega
\pm \omega_J|^{2g-1}$. Measurements of the location of this
singularity would give the value $e^*$ of  the charge of the
carriers of the current, which would be yet another way of
observing fractional charge from noise measurements. The method
originally suggested is to measure the shot noise, which for small
tunneling amplitude is related to the tunneling current $I_t$ by
$S_{SN}=2e^* I_t$ \cite{Kane&Fisher2,CFW,Fendley3}. Lastly, for $g
> 1$, we found that the singularities at both $\omega = 0$ and
$\omega = \omega_J$ should persist to all orders in perturbation
theory.

These results present a puzzle which we describe below and address in
this paper. The case of $g=\nu <1$ corresponds to a single quantum
Hall droplet with a constriction. In this case, quasiparticles can
tunnel across the constriction, from one edge to the other (see
Fig. 1a).  These quasiparticles have fractional charge $e^*$, given
by $\nu e$. If the constriction is made narrower, the tunneling
amplitude will increase.  As the constriction is further narrowed,
eventually the droplet will break into two disconnected pieces, and
now only electrons will be able to tunnel from one edge to the
other (see Fig. 1b). Their tunneling should once again behave like
tunneling in a chiral Luttinger liquid, but with new exponent
$\tilde g = 1/g$ and charge equal to $e$.  This is the physical
picture behind the duality seen in reference \cite{Fendley2}; as
the tunneling amplitude is increased (or the voltage is decreased)
$g$ goes to $1/g$. Similarly, if we start with the two quantum Hall
droplets with exponent $\tilde g$ and increase the
tunneling amplitude of the electron, eventually we will obtain the
single droplet picture with exponent $g = 1/\tilde g$.

In light of this duality and the results of reference \cite{CFW}, the
following question arises.  If we start with the two disconnected
droplets then we expect the singularity in the noise to occur at
multiples of $\tilde\omega_J = eV/\hbar$, the Josephson frequency for
the electron.  As the tunneling amplitude is increased, at some point
we expect the singularity at the Josephson frequency for the
quasiparticle, $\omega_J = e^*V/\hbar$, to appear. However, according
to the perturbative calculations, to all orders in the electron
tunneling amplitude the quasiparticle singularity does not appear. In
this paper we will address the question of what happens to this
quasiparticle singularity and show how these two seemingly
contradictory statements are resolved in the special case of $g = 1/2$
and $\tilde g = 2$. In short, our results suggest that the singularity
at the quasiparticle Josephson frequency $\omega_J=e^*V/\hbar$ is
destroyed by non-perturbtive effects. In some sense, the singularity
obtained by perturbative calculations is smeared for finite tunneling
strength. This question is of interest because the location of the
singularities tells us which particles are tunneling and, as mentioned
above, should give a way to measure the fractional charge of the
quasiparticles.  The results in this paper suggest what may be
happening at other values of $g$ also, and we are currently working on
this question.

One of the tools we use in this paper is the Landauer-Buttiker
scattering approach. The geometry is illustrated in Figure 2. The
choice of the Landauer-Buttiker approach is justified for a number
of reasons. The chiral nature of the system under study naturally
poses the problem in terms of incoming and outgoing scattering
states to and from the point contact region. The incoming branches
should be in equilibrium with their respective reservoirs of
departure, and should be insensitive to the tunneling of charges in
the tunneling region shown in Fig. 2. This is so because
information on tunneling events cannot propagate in the direction
opposite to the incoming branch chirality. Also, the Landauer-Buttiker
approach and the chiral nature of the system suggest
naturally a four-terminal geometry for experimental measurements,
probing voltage fluctuations in the two incoming and two outgoing
branchs. The tunneling takes place in the point contact, or
scattering region, which is not directly accessible by the probing
leads. Auto-correlations of current/voltage fluctuations measured
in the four terminals, as well as cross-correlations between
different terminals, are the experimental probes that should allow the
remote measurement of the tunneling events and noise spectrum.

The paper is organized as follows. In section II we briefly review the
bosonization scheme for chiral Luttinger liquids. In section III we
obtain the noise spectrum perturbatively for the four terminal
geometry, using the Keldysh non-equilibrium formalism. We show that
only the noise spectrum for the outgoing branches is affected by the
tunneling, whereas the incoming branches are completely insensitive to
the charge transfer between the edges. This is consistent with the
Landauer-Buttiker picture and the chirality of the system. The noise
spectrum for the incoming branches can thus serve as reference level
for the measurement of the excess noise on the outgoing branches due
to tunneling. The noise spectrum obtained contains interesting
structures both at low and high frequencies.  The tunneling excess
noise vanishes for frequencies above the Josephson frequency
$\omega_J=e^*V/\hbar$. The issue of how the singularity moves from the
quasiparticle frequency to the electron frequency is resolved in
section IV, where we use the Landauer-Buttiker approach to solve
exactly for the noise spectrum in the case of $g=1/2$, for which the
problem can be cast as a free fermion problem. We show that the
singularity at the quasiparticle frequency is smeared for finite
tunneling and is not a true singularity, whereas the singularity at
the electron frequency survives for all non-zero coupling. In section
V we discuss the duality when $g=1/2$ goes to ${\tilde g}=2$, which we
show is not exact in the naive sense for the case of noise, in
contrast to the case of conductance.  We find that the noise spectrum
of the current correlations on a single branch (auto-correlations)
satisfies the duality relation, while current correlations between
distinct branches (cross-correlations) do not satisfy the naive
duality relation. We show that the correct dual Lagrangian to the
$g=1/2$ theory is the $g=2$ theory plus a neutral density-density
coupling, which has the same dimension as the tunneling operator. The
effect of the neutral coupling appears in the noise, but not in the
conductance.

\section{Edge State Tunneling}

In this section we shall briefly review the bosonization scheme for
edge states in the FQH effect (for a thorough review, see Ref.
\cite{XGWreview}). The Lagrangian we will use is better cast in
this bosonic language.

The right and left moving excitations along the edges can be described
by boson fields $\phi_{R,L}$. Right and left moving electron and
quasiparticle operators on the edges of a FQH liquid can be written as
$\Psi_{R,L}(t,x)\propto e^{\pm i \sqrt{g} \phi_{R,L} (t,x)}\ $, where
$g$ is related to the FQH bulk state. For example, for a Laughlin
state with filling fraction $\nu=1/m$ we have $g=m$ for electrons and
$g=1/m$ for quasiparticles carrying fractional charge $e/m$. The
$\phi_{R,L}$ fields satisfy the equal-time commutation relations
\begin{equation}
[\phi_{R,L}(t,x)\ ,\ \phi_{R,L}(t,y)]=\pm i\pi \ {\rm sign}(x-y)\
{}.
\end{equation}
The dynamics of $\phi_{R,L}$ is described by
\begin{equation}
{\cal L}_{R,L}=\frac{1}{4\pi}\ \partial_x\phi_{R,L}
\ (\pm \partial_t- v \partial_x)\phi_{R,L}\ ,
\end{equation}
where $v$ is the velocity of edge excitations (which we will set to
1).  Density operators can be defined in terms of the $\phi_{R,L}$
through $\rho_{R,L}=\frac{\sqrt{\nu}}{2\pi}\partial_x\phi_{R,L}$.
Here, for convenience, we have set the unit charge in the
definition of the density to be the electron charge $e$, so that
$e=1$ and $e^*=\nu$. One can verify that $[\rho_{R,L}(t,x)\ ,\
\Psi^\dagger_{R,L}(t,y)]=\sqrt{\nu g}\
\Psi^\dagger_{R,L}(t,y)\delta (x-y)$, so that indeed the cases
$g=\nu^{-1}$ and $g=\nu$ correspond to electron and quasiparticle
charged operators, respectively.

The tunneling operators from right to left moving branches and
vice-versa can be written as $\Psi^\dagger_L \Psi_R$ and
$\Psi^\dagger_R \Psi_L$. Thus we can write, in terms of
$\phi=\phi_R + \phi_L$, the following total Lagrangian density:
\begin{equation}
{\cal L}=\frac{1}{8\pi}[(\partial_t\phi)^2-v^2
(\partial_x\phi)^2]-
\ \Gamma \delta(x)\ e^{i\sqrt{g}\phi(t,0)}+H.c.\ \ ,\label{L}
\end{equation}
with $\phi$ satisfying $[\phi(t,x),\partial_t\phi(t,y)]=4\pi
i\delta(x-y)$. In the following we will set the edge velocity $v=1$.

A voltage difference between the two edges of the QH liquid can be
easily introduced in the model by letting $\Gamma \rightarrow
\Gamma e^{-i\omega_0 t}$, where $\omega_0\equiv \omega_J \equiv
e^*V/\hbar$, with $e^*=e$ for electron tunneling and $e^*=e/m$ for
quasiparticle tunneling.

In the following sections we will study non-equilibrium noise in
chiral Luttinger liquids described by the model above.

\section{Perturbative Approach}

In this section we treat the tunneling between edge states
perturbatively, and obtain the noise spectrum for the
current/voltage fluctuations at the four leads as shown in Fig. 2.
In the figure we separate the branches into their right and left
moving components, as well as incident and scattered ones. Right
and left branches are incoming or outgoing depending on their
position relative to the scatterer:
\begin{eqnarray}
{\rm Incident}\ \ &\ &\ \phi_R(t,x<0)\ \ {\rm and}\ \ \
\phi_L(t,x>0)\nonumber \\ {\rm Scattered} \ &\ &\ \phi_R(t,x>0)\ \
{\rm and}\ \ \ \phi_L(t,x<0)\ .
\end{eqnarray}
Both the currents and the densities at the four terminals can be
related to the fields $\phi_{R,L}$. The densities are simply given
by $\rho_{R,L}=\frac{\sqrt{\nu}}{2\pi}\partial_x\phi_{R,L}$.
Voltage measurements probe these densities. The currents at the
four terminals can be trivially related to the densities at those
same terminals through the continuity equation for $x\neq 0$. The
currents are given by
$j_{R,L}=\pm\frac{\sqrt{\nu}}{2\pi}\partial_x\phi_{R,L}$, with
positive currents flowing to the right. By choosing the convention
that positive currents flow in the direction of the arrows in Fig.
2, we can write new currents ${\tilde j_{R,L}}=\pm
j_{R,L}=\rho_{R,L}$. It then becomes transparent that there is a
tight relationship between current and voltage in the chiral
branches. For example, measuring the noise in either the current
or the voltage yields information about the other.  This kind of
relationship between voltage and current noise was obtained in Ref.
\cite {Kane&Fisher2}. We will thus focus on the calculation of
density-density correlations, for these will give us information on
both current and voltage noise.

We will label the densities at the four terminals shown in Fig. 2
by $\rho_i$, $i=1,2,3,4$. In terms of the right and left moving
fields we have:
\begin{eqnarray}
\rho_1(t)=\rho_R(t,x_1)\ \ &\ &\ \
\rho_3(t)=\rho_R(t,x_3)\nonumber\\ \rho_2(t)=\rho_L(t,x_2)\ \ &\ &\
\ \rho_4(t)=\rho_L(t,x_4)\ ,
\label{rhonumdef}
\end{eqnarray}
where $x_1,x_4<0$, $x_2,x_3>0$.
The noise spectrum of the density fluctuations in terminals $i,j$
is obtained from the correlations between the densities
$\rho_i,\rho_j$: \begin{equation}
S_{ij}(\omega)=S_{ji}(-\omega)=\int_{-\infty}^\infty dt\ e^{i\omega
t} \langle \{\rho_i(t),\rho_j(0)\}\rangle\ .
\end{equation}
These quantities are calculated perturbatively in Appendix A, using
the techniques in Ref. \cite{CFW}. The components with $i\neq j$
are very sensitive to phases which depend on the position of the
probes $x_i$ and $x_j$.  These phases cancel in the case of
auto-correlations, {\it i.e.}, when $i=j$. The quantities
$S_{ii}(\omega)$, which correspond to the noise spectrum obtained
entirely from one of the four probes for $i$=1 to 4, are thus the
most robust measurements of fluctuations, because when they are
extracted away from the junction they are independent of the
position $x_i$ where they are taken.

To second order in perturbation theory, $S_{ii}$ is given by:
\begin{eqnarray}
S_{11}(\omega)&=&S_{22}(\omega)=S^{(0)}(\omega)\label{s11secIII}\\
S_{33}(\omega)&=&S_{44}(\omega)=S^{(0)}(\omega)+S^{(2)}(\omega)\
,\label{s33secIII}
\end{eqnarray}
where
\begin{eqnarray}
S^{(0)}(\omega)&=&\frac{\nu}{2\pi}|\omega|\ ,\label{s0secIII}\\
S^{(2)}(\omega)&=&\frac{4\pi\nu g}{\Gamma(2g)}\ |\Gamma|^2\ \
\Big||\omega| -|\omega_J|\Big|^{2g-1}\ \theta(|\omega_J|-|\omega|)\
.\label{s2secIII}
\end{eqnarray}
Using the perturbative result to order $|\Gamma|^2$ for the
tunneling current
$I_t=\frac{2\pi}{\Gamma(2g)}e^*|\Gamma|^2\omega_J^{2g-1}$
\cite{XGW2}, $S^{(2)}(\omega)$ can be wrtitten as
\begin{equation}
S^{(2)}(\omega)=2e^*I_t \
\Big|1-\Big|\frac{\omega}{\omega_J}\Big|\Big|^{2g-1}\
\theta(|\omega_J|-|\omega|)\ .\label{noise2ndorder}
\end{equation}
Notice that the effects of tunneling are contained in
$S^{(2)}(\omega)$, and only appear in the outgoing branches,
terminals $i=3,4$. The incoming branches are insensitive to the
tunneling between edges, due to the chiral nature of the system.
Information about what goes on in the junction cannot propagate in
the direction opposite to the chirality of the branch, and
therefore the noise in the incoming branches is independent of the
tunneling of charged particles between edges. This result of
chirality is clear within the Landauer-Buttiker scattering
approach. Another physical consequence closely related to this is
the fact that the average voltage along the branches remains
constant outside the scattering region.

The second point to notice from Eq. (\ref{noise2ndorder}) is that to
order $|\Gamma|^2$ the noise in the outgoing branches that is in
excess to the noise in the incoming branches has a singularity at the
Josephson frequency $\omega_J$, vanishing for $\omega>\omega_J$, as
illustrated in Fig. 3. The non-equilibrium voltage $V$ determines the
frequency scale $\omega_J=e^*V/\hbar$, up to which there is structure
in the excess noise due to tunneling. Such vanishing of the excess
noise spectrum past a frequency set by the non-equilibrium voltage
should be familiar to readers accustomed to noise in non-interacting
systems ($g=1$), in which the excess noise goes to zero linearly at
the Josephson frequency \cite{Yang}. This point will be illustrated
further in the next section, when we will have at hand the exact
solution for the noise spectrum in the case of $g=1/2$.  The strong
coupling limit of the solution for $g=1/2$ also gives us the solution
for $g=2$, which we shall use for comparison purposes.

The last, and most important, point about this high frequency
singularity in the noise spectrum is in regard to the connection
between the two dual pictures illustrated in Fig. 1. In Ref.
\cite{CFW} it was pointed out that the singularity at the Josephson
frequency remained to all orders in perturbation theory. However,
the perturbative expansion for the geometries in Figs. 1a and
1b yields two distinct frequencies, namely the quasiparticle
frequency $\omega_{qp}=\nu eV/\hbar$ when quasiparticles are the
tunneling charges (Fig. 1a), and the electron frequency
$\omega_{el}=eV/\hbar$, when electrons are the tunneling current
carriers (Fig. 1b). These configurations are connected in the sense
that one is the strong coupling limit of the other, and thus there
should be a
non-perturbative mechanism by which the singularity moves from one
place to the other.  This was the clearest open question in Ref.
\cite{CFW}, and which we can answer by focusing on the exactly
solvable case of $g=1/2$.  Another exactly solvable point is the
trivial case $g=1$, which unfortunately cannot be used to address
this issue of the singularity in the noise spectrum because in this
case the two frequencies $\omega_{el}$ and $\omega_{qp}$ coincide.

Before answering the question about the high frequency singularity,
we will close this section with the implications of tunneling
between edge states to the low frequency noise measured in the four
terminal geometry. In Ref. \cite{CFW}, a correction to the low
frequency shot noise spectrum was found, which corresponded to an
$|\omega|$ singularity, or a cusp, in the noise spectrum. This
correction was found to order $|\Gamma|^4$, while to order
$|\Gamma|^2$ the low frequency corrections to the flat shot noise
started as $\propto \omega^2$. In the four terminal
geometry proposed in this paper, what is probed is not the
tunneling current in the junction area (as in Ref. \cite{CFW}), but
its consequences in the
current/voltage in the four terminals away from the scattering
region. The four terminal measurement, as seen from Eq.
(\ref{noise2ndorder}), does have a correction $\propto |\omega|$ to
order $|\Gamma|^2$. For $\omega\ll\omega_J$ we have, for example,
\begin{eqnarray}
S_{33}(\omega)-S^{V=0}_{33}(\omega)&=&S^{(2)}(\omega)=
S_{33}(\omega)-S_{11}(\omega)\nonumber\\
&=&2e^*I_t \
\Big|1-\Big|\frac{\omega}{\omega_J}\Big|\Big|^{2g-1}\
\theta(|\omega_J|-|\omega|)\\
&\approx&2e^*I_t \ \left[1-
(2g-1)\Big|\frac{\omega}{\omega_J}\Big|\right]\ . \nonumber
\end{eqnarray}
One recovers the classical shot noise expression for $\omega=0$.
Notice that, since these results are valid only to order $|\Gamma|^2$,
there is no correction to the classical shot noise expression for
$\omega=0$. Corrections appear at order $|\Gamma|^4$ (see Ref.
\cite{CFW}). Also notice that the non-zero $\omega$ corrections to
the shot noise depend on whether $g$ is larger or smaller then
$1/2$. For
$g>1/2$, the difference between the outgoing and incoming spectra
(the $S_{33}(\omega)-S_{11}(\omega)$ above, for example) decreases
with $\omega$, whereas for $g<1/2$ it increases.


\section{Scattering Approach for \hbox{$g=1/2$}}
In this section we will use the Landauer-Buttiker Scattering
approach to obtain an exact solution for the noise when $g = 1/2$.
In this approach, we use the quantum equations of motion derived
from the Hamiltonian to solve for the scattering states.  These
scattering states describe free left movers and right movers that
are incident on the impurity and then are reflected or scattered by
the impurity.  The solutions for these states can be used to
calculate the conductance and the noise in the various branches.

The advantage of focussing on $g = 1/2$ is that for this value of $g$
the system can be described by free fermions \cite{Guinea2,Matveev},
making it straightforward to solve for the scattering states.
However, already at $g = 1/2$, we expect to see singularities in the
noise at $\omega_J = e^*V/\hbar$, corresponding to quasiparticle
tunneling. As the tunneling amplitude $\Gamma$ increases (or $V$
decreases), we expect to obtain the dual picture at $g = 2$, with
electrons tunneling and a singularity at $eV/\hbar$.  Thus the full
solution at $g = 1/2$ will show us what happens to the quasiparticle
singularity as $\Gamma$ is increased.  The hope is that the
qualitative behavior of these results will also apply for other values
of $g$.

When $g= 1/2$, the Hamiltonian for the system is given by
\begin{equation}
H = H^0_R + H^0_L + \Gamma e^{-i\omega_0 t}\
e^{\frac{i}{\sqrt{2}}(\phi_R(t,0)+\phi_L(t,0))} +  \Gamma^*
e^{i\omega_0 t}\ e^{-\frac{i}{\sqrt{2}}(\phi_R(t,0)+\phi_L(t,0))}
\end{equation}
where $H^0_{R,L}$ are the free Hamiltonians for the right and left
moving fields, and $\omega_0=e^*V/\hbar$, with $e^*=e/2$.

The Hamiltonian can be recast in terms of new chiral
fields $\phi_{\mp}(t,x)=\frac{1}{\sqrt{2}}(\phi_R(t,x)\pm
\phi_L(t,-x))$:
\begin{equation}
H = H^0_+ + H^0_- + \Gamma e^{-i\omega_0 t}\ e^{i\phi_-(t,0)} +
\Gamma^* e^{i\omega_0 t}\ e^{-i\phi_-(t,0)}\ . \label{Hpm}
\end{equation}
The densities of the new fields
$\rho_\pm=\frac{1}{2\pi}\partial_x\phi_\pm$ are related to the
densities $\rho_{R,L}=\frac{\sqrt{1/2}}{2\pi}\partial_x\phi_{R,L}$
by $\rho_\pm(t,x)=\rho_R(t,x)\pm\rho_L(t,-x)$. Notice that the
$\phi_\pm$ fields are decoupled in Eq.(\ref{Hpm}), and the
Hamiltonian for $\phi_+$ is simply the free $H^0_+$. The
Hamiltonian for $\phi_-$ can be fermionized by defining
$\eta(t,x)\equiv
\frac{1}{\sqrt{2\pi}}:e^{i\phi_-(t,x)}:\ $. One can check that
$\eta$ defined as such satisfies the proper commutation relations
$\{\eta(t,x)\ ,\ \eta^\dagger(t,y)\}=\delta(x-y)$
\cite{Floreanini&Jackiw}.

In terms of the fermionic fields $\eta,\eta^\dagger$, the
Hamiltonian $H_-$ is:
\begin{equation}
     H_- = \int dx\
 {\Big \{}
\eta^\dagger(x)\left[-i{\partial \over\partial x} - \omega_0\right]
\eta(x)    + \sqrt{2\pi}\delta(x)\left[\Gamma \eta(x) + \Gamma^*
\eta^\dagger(x)\right] {\Big \}},    \label{Heta}
\end{equation}
where we absorbed the oscillating phases $e^{i\omega_0 t}$ into a
redefinition of the chemical potential. The Hamiltonian above
contains terms linear in the fermionic fields $\eta$ and
$\eta^\dagger$, which prevent a direct calculation of the
commutators that would give us the equations of motion for the
fields. This problem can be circumvented by redefining the
fermionic fields to be
$\psi(t,x)=\eta(t,x) f$, with $f=C+C^\dagger$ and
$\{C,C^\dagger\}=1$, as in Ref. \cite{Matveev}. More formally, such
a transformation can be constructed from the proper handling of the
zero modes of the bosonic fields $\phi$ \cite{XGWcll,GSW}, and
one can identify $f$ with $(-1)^F$, the fermion counting operator
commonly used to switch from periodic to anti-periodic boundary
conditions in fermionic conformal field theories.

The Hamiltonian we will use in the exact solution of the noise
spectrum for the $g=1/2$ case is the one written in terms of the
$\psi,\psi^\dagger$ fields and $f$:
\begin{equation}
     H_- = \int dx\
{\Big \{}
\psi^\dagger(x)\left[-i{\partial \over\partial x} - \omega_0\right]
\psi(x)    + \sqrt{2\pi}\delta(x)\left[\Gamma \psi(x)f + \Gamma^*
f \psi^\dagger(x) \right]  {\Big \}}, \label{Hpsi}
\end{equation}
where the non-vanishing equal-time
commutation relations
between $\psi(x),\psi^\dagger(x)$ and $f$ are
\begin{equation}
\{\psi(x), \psi^\dagger(x')\}$=$ \delta(x - x')\ ,\
\{\psi(x), f\}=0\ ,\ \{f,f\}=2\ . \label{psicom}
\end{equation}

The density $\rho_-$ can be written in terms of the fields
$\psi$ and $\psi^\dagger$ as $\rho_-(x)=\psi^\dagger(x)\psi(x)$, so that
all correlations between $\rho_-$'s can be derived from the
correlations of the fermions. The fermionic model is solved using
the equations of motion obtained by commuting the operators
$\psi(x)$ and $f$ with the Hamiltonian:
\begin{eqnarray}
-i\partial_t \psi(x) &=& [H, \psi(x)]
                     = (i\partial_x + \omega_0) \psi(x) +
\sqrt{2\pi}\Gamma^* f\delta(x),    \label{eompsi}\\
-i\partial_t \psi^\dagger(x) &=& [H, \psi^\dagger(x)]
    = (i\partial_x - \omega_0) \psi^\dagger(x) -
\sqrt{2\pi}\Gamma f\delta(x),    \label{eompsid}
\end{eqnarray}
and
\begin{equation}
-i\partial_t f = [H, f]
              = 2\sqrt{2\pi}\left[\Gamma \psi(0) -
\Gamma^*\psi^\dagger(0)\right].    \label{eomf}
\end{equation}
According to these equations, for $x \ne 0$, the field $\psi$
satisfies the free equation of motion for a rightmover with energy
shifted by $\omega_0$:
\begin{equation}
(i\partial_x + i\partial_t + \omega_0)\psi = 0. \label{eomrm}
\end{equation}
At $x = 0$, it picks up a discontinuity because of the impurity.
In order to  preserve unitarity and obtain the proper commutation
relations in the solutions of $\psi$, in equation (\ref{eomf}) the
field $\psi(0)$ must be given by  $(1/2)(\psi(0^+) + \psi(0^-))$.
With this definition, it is straightforward to solve the equations
of motion.  The solutions are given by
\begin{equation}
\psi(x) = \cases{\sum_\omega A_\omega e^{i(\omega + \omega_0)x}
e^{-i\omega t}                       & for $x < 0$
\label{psixminus}\cr
                 \sum_\omega B_\omega e^{i(\omega + \omega_0)x}
e^{-i\omega t}                       & for $x > 0$
\label{psixplus}\cr}
\end{equation}
and
\begin{equation}
\psi^\dagger(x) = \cases
     {\sum_\omega A^\dagger_{-\omega} e^{i(\omega - \omega_0)x}
e^{-i\omega t}                       & for $x < 0$
\label{psidxminus}\cr
     \sum_\omega B^\dagger_{-\omega} e^{i(\omega - \omega_0)x} e^{-
i\omega t}                       & for $x > 0$,
\label{psidxplus}\cr}  \end{equation}
where
\begin{equation}
B_\omega = {(1 + e^{i\phi(\omega)})A_\omega
           + (1-e^{i\phi(\omega)})A^\dagger_{-\omega} \over 2},
\label{Bdef}
\end{equation}
and
\begin{equation}
e^{i\phi(\omega)} = {i\omega + 4\pi|\Gamma|^2 \over i\omega -
4\pi|\Gamma|^2 }. \label{phidef}
\end{equation}
Given the commutation relation for $\psi$, the $A_\omega$ satisfy
the  following commutation relation:
\begin{equation}
\{A_{\omega_1}, A^\dagger_{\omega_2}\} = \delta_{\omega_1,
\omega_2}. \label{Acom}
\end{equation}
These solutions can be interpreted as having an incident particle
at energy $\omega$ that scatters into a particle with energy
$\omega$ and a hole with energy $-\omega$ (see Fig. 4). Both the
particle and hole scattering involve an energy dependent phase
shift.

The reservoir is located to the left of the impurity, for some $x
< 0$.   To obtain the scattering state $|\Phi\rangle$, we assume
that the states  leaving the reservoir are in equilibrium with the
reservoir, which has energy $\omega_0$.  Thus, for $x < 0$, at zero
temperature all the states with  $\omega \le \omega_0$ are filled.
This means that
\begin{equation}
A^\dagger_\omega |\Phi \rangle = 0 \qquad {\rm for} \quad \omega <
\omega_0, \label{AdPhidef}
\end{equation}
and
\begin{equation}
A_\omega |\Phi \rangle = 0 \qquad {\rm for} \quad \omega >
\omega_0. \label{APhidef}
\end{equation}
Using the commutation relations for $A$ in equation (\ref{Acom}),
we then find that
\begin{equation}
 \langle \Phi | A_{\omega_1} A_{\omega_2} |\Phi \rangle = 0,
\label{AAexp}
\end{equation}
and
\begin{equation}
 \langle \Phi | A^\dagger_{\omega_1} A_{\omega_2} |\Phi \rangle
  = n_{\omega_1}\delta_{\omega_1, \omega_2},
\label{AdAexp}
\end{equation}
where
\begin{equation}
   n_\omega = \cases{1 & for $\omega < \omega_0$ \cr
                     0 & for $\omega > \omega_0$. \cr}
\label{nzT}
\end{equation}
In this paper, we will just concentrate on the case when $T=0$.
However, we can obtain the finite temperature results by replacing
$n_\omega$ with
\begin{equation}
   n_\omega = {1\over e^{\beta(\omega-\omega_0)} + 1}.
\label{nfiniteT}
\end{equation}

We can now use the solutions for $\psi$ and the scattering state to
solve for the noise in both incoming and outgoing channels. Our
calculations will closely follow those by Buttiker in reference
\cite{Buttiker}. The noise is given by
\begin{equation}
S(\omega; x_1, x_2) = \int_{-\infty}^\infty dt\ e^{i\omega t}
       \langle \{\rho_-(t, x_1), \rho_-(0, x_2)\} \rangle,
\label{Sijdef}
\end{equation}
where we take only the connected part of the correlation function,
and $x_1$ and $x_2$ are positive or negative depending on whether
the current is evaluated in the incoming or outgoing channel.

We will begin by calculating the noise when $x_1=x_2$. In this case,
both of the currents are evaluated on the same side of the impurity.
Because of the time translational invariance of the correlators, the
expression for the noise simplifies to
\begin{equation}
S(\omega; x_1, x_1) = \int_{-\infty}^\infty dt
      \left(e^{i\omega t} + e^{-i\omega t}\right)
            \langle \rho_-(t, x_1) \rho_-(0, x_1) \rangle.
\label{Sjjdef}
\end{equation}
To find the noise in the incoming channel, we must evaluate the
expectation value
\begin{equation}
\langle \rho_-(t, x_-) \rho_-(0, x_-) \rangle
  = \langle \psi^\dagger(t,x_-)\psi(t, x_-)\psi^\dagger(0,x_-
)\psi(0,x_-)      \rangle,
\label{ImImdef}
\end{equation}
with $x_- < 0$.  Using the solutions (\ref{psixminus}) and
(\ref{psidxminus}) for $\psi$ and $\psi^\dagger$, we find
\begin{equation}
\langle \rho_-(t, x_-) \rho_-(0, x_-) \rangle
  = \sum_{\omega_1,\omega_2,\omega_3,\omega_4} e^{-
i(\omega_1+\omega_2)t}      \langle \Phi |
     A_{-\omega_1}^\dagger A_{\omega_2} A_{-\omega_3}^\dagger
A_{\omega_4}      | \Phi \rangle
      e^{i(\omega_1 + \omega_2 + \omega_3 + \omega_4) x_-}.
\label{ImImcalc}
\end{equation}
This expectation value, and the resulting integrals for
$S(\omega;x_-,x_-)$ are evaluated in Appendix C, with the result,
\begin{equation}
S(\omega;x_-,x_-) = \frac{1}{2\pi}|\omega|.
\label{Smmans}
\end{equation}

If we want to calculate the noise in one of the two original $R$
and $L$ incoming branches, we must use the relations
\begin{equation}
\rho_R(x) = {1\over 2} (\rho_+(x) + \rho_-(x)) \qquad {\rm and}
\qquad \rho_L(x) = {1\over 2} (\rho_+(-x) - \rho_-(-x)).
\label{rhoIIdef}
\end{equation}
Then the density-density correlations can be evaluated as follows:
\begin{equation}
\langle \rho_{R,L} \rho_{R,L} \rangle
     = {1\over 4} \langle (\rho_+ \pm \rho_-)(\rho_+ \pm \rho_-)
\rangle      = {1\over 4} \langle \rho_- \rho_- \rangle + {1\over
4} \langle \rho_+ \rho_+ \rangle, \label{rhorhoII}
\end{equation}
where the last equality follows from the fact that $\rho_+$ and
$\rho_-$ are decoupled. Recall that $\rho_+$ is a free field, so
that the contribution to the noise from $\rho_+$ is simply
$\frac{1}{2\pi}|\omega|$. We find that the noise in each of the two
incoming $R$ and $L$ branches is given by
\begin{eqnarray}
S_{11}(\omega) = S_{22}(\omega) &=& {1\over 4} S(\omega;x_-,x_-)
  + {1\over 4} \frac{|\omega|}{2\pi}\label{Soneone}\\
&=&{1\over 4\pi} |\omega|,\nonumber
\end{eqnarray}
just as we found in the perturbative calculation with $\nu=1/2$ in
Eqs. (\ref{s11secIII}) to (\ref{s2secIII}). Using this scattering
approach, it is clear that for these two incoming probes the noise
is the same as for a free
system, because in these two channels the densities have not yet
reached the impurity.

Next, we will calculate the noise in the outgoing current.  This
time we must evaluate the correlator
\begin{equation}
\langle \rho_-(t, x_+) \rho_-(0, x_+) \rangle
  = \langle \psi^\dagger(t,x_+)\psi(t,
x_+)\psi^\dagger(0,x_+)\psi(0,x_+)      \rangle,
\label{IpIpdef}
\end{equation}
with $x_+ > 0$.  According to equations (\ref{psixplus}) and
(\ref{psidxplus}), this is equal to
\begin{equation}
\langle \rho_-(t, x_+) \rho_-(0, x_+) \rangle
  = \sum_{\omega_1,\omega_2,\omega_3,\omega_4} e^{-
i(\omega_1+\omega_2)t}      \langle \Phi |
     B_{-\omega_1}^\dagger B_{\omega_2} B_{-\omega_3}^\dagger
B_{\omega_4}      | \Phi \rangle
      e^{i(\omega_1 + \omega_2 + \omega_3 + \omega_4) x_+}.
\label{IpIpcalc}
\end{equation}

When we expand the $B$'s in terms of the $A$'s, we will obtain two
different types of processes (see Fig. 5).  In the first, at time
$0$ one particle is created while another is destroyed, and then at
time $t$ the first particle is destroyed and another is created.
In terms of the original tunneling picture, this describes the
process where both at time $t$ and at time $0$ one quasiparticle
tunnels from the left branch to the right branch and another
tunnels in the opposite direction.  In the second process, at time
$0$ two particles are created and then at time $t$ they are
destroyed (or {\it vice versa}). In the original tunneling picture,
this corresponds to two quasiparticles tunneling in one direction
at time $0$ and two quasiparticles tunneling in the opposite
direction at time $t$.  As shown in Appendix C, this second process
is responsible for the electron singularity at  $\tilde\omega_0 =
2\omega_0$.
In Appendix C, the expectation values in Eq.(\ref{IpIpcalc}) and
the integrals for $S(\omega;x_+,x_+)$ are evaluated.
We find that the
noise on the outgoing side of the impurity is
\begin{eqnarray}
S(\omega;x_+,x_+)(\omega) = \frac{1}{2\pi}|\omega| +
&\theta&(|2\omega_0| - |\omega|)
\bigg\{4|\Gamma|^2\left[\tan^{-1}\left({|\omega_0|\over
4\pi|\Gamma|^2}\right)             +\tan^{-1}\left({|\omega_0| -
|\omega|\over 4\pi|\Gamma|^2}\right)          \right]  \nonumber\\
    &+&{16\pi|\Gamma|^4\over |\omega|}
     \left[\ln\left((4\pi|\Gamma|^2)^2+(|\omega| -
|\omega_0|)^2\right)             -\ln\left((4\pi|\Gamma|^2)^2
+\omega_0^2\right)\right]       \bigg\}.
\label{Spp}
\end{eqnarray}
To compare with our perturbative calculation for the noise in the
original four probe geometry, we again make use of equation
(\ref{rhorhoII}).  Thus, the noise in the two outgoing branches is
related to $S(\omega;x_+,x_+)$ as follows:
\begin{equation}
S_{33}(\omega) = S_{44}(\omega) = {1\over 4} S(\omega;x_+,x_+)
+ {1\over 4} \frac{|\omega|}{2\pi}.
\end{equation}

The first striking feature to note in equation (\ref{Spp}) is that
the noise due to the tunneling vanishes identically for $|\omega|
> |2 \omega_0|$.  This means that whenever $|\omega|$ is larger
than the electron frequency, the noise shows no sign of the
impurity; it is the same as for the incoming branch.  This is also
what happens for the free electron case, with $g=1$. To second
order in perturbation theory, this is indeed the case for any $g$,
as seen in the previous section. The strength of the results
presented here is that for {\it any value} of the coupling $\Gamma$
the noise vanishes above the electron frequency when $g=1/2,1$ and
$2$.  (The last case, $g=2$, is obtained by resorting to the strong
coupling limit of the $g=1/2$ case.) It is not clear whether this
will happen for the other values of $g$ beyond second order in
perturbation theory.

Next, we can expand $S(\omega;x_+,x_+)$ out for small and large
$|\Gamma|$ to compare with the perturbative results.  As $|\Gamma|$
goes to zero, the noise becomes
\begin{equation}
S_{33}(\omega) = S_{44}(\omega) = {1\over 4\pi}|\omega|
                 + \pi |\Gamma|^2 \theta(|\omega_0| - |\omega|).
\label{SttsmallG}
\end{equation}
This agrees with the perturbative result for $g = 1/2$.  We note
that the quasiparticle singularity arises because we took the
$|\Gamma| \to 0$ limit of the arctangents.  In addition, because
this step function is already zero for $|\omega| > |\omega_0|$, the
electron singularity at
$|\omega| = |2\omega_0|$ drops out.  Thus, to this order we only
have the quasiparticle singularity.  However, for any finite value
of $|\Gamma|$ the quasiparticle singularity becomes smoothed out
and the electron singularity appears.  As we shall see later,
though,  the ``smoothed out'' quasiparticle singularity is still a
more distinctive feature in the plots of the full noise than is the
electron singularity.

Next, for $|\Gamma| \to \infty$, the noise becomes
\begin{equation}
S_{33}(\omega) = S_{44}(\omega) = {1\over 4\pi}|\omega| +
    {1\over 384\pi^3|\Gamma|^4} \theta\left(|2\omega_0| -
|\omega|\right)      \left(|2\omega_0| - |\omega|\right)^3
+O(1/|\Gamma|^8).
\label{SttscatlG}
\end{equation}
If we make the identification that $\Gamma_{1\over2}$, the
tunneling amplitude for $g = 1/2$, is related to $\Gamma_2$, the
tunneling amplitude for $g=2$, by
\begin{equation}
|\Gamma_2| = {1\over{16\pi^2|\Gamma_{1\over2}|}^2},
\label{GtGoh}
\end{equation}
then this answer agrees with the perturbative result for $g=2$.
(To make the comparison, we must recall that the $\omega_0$ in this
equation corresponds to the Josephson frequency for the
quasiparticle, whereas the $\omega_J$ in the perturbative
calculation Eqs. (\ref{s11secIII}) to (\ref{s2secIII}) is the
Josephson frequency for the
electron, which is twice as large.) In addition, the expansion in
${1\over |\Gamma|}$ of the scattering solution only contains powers
of ${1\over |\Gamma|^4} = |\Gamma_2|^2$, and at every order in
$|\Gamma_2|^2$ the electron singularity at $|\omega| = |2\omega_0|$
remains.  These two properties also agree with the perturbative
results found in
\cite{CFW}.

Lastly, we can make use of the scaling properties of the noise to
write $\tilde S = S/2|\Gamma|^2$ as a function only of
$\tilde \omega = {\omega\over 4\pi|\Gamma|^2}$ and
$\tilde \omega_0 = {\omega_0\over 4\pi|\Gamma|^2}$. The noise is
then given by
\begin{eqnarray}
\tilde S_{33}(\tilde \omega) &=& \tilde S_{44}(\tilde \omega)
\nonumber\\    &=&
{1\over 2}|\tilde \omega| + \theta(|2\tilde \omega_0| - |\tilde
\omega|)     \bigg\{{1\over
2}\left[\tan^{-1}\left(|\tilde\omega_0|\right)
+\tan^{-1}\left(|\tilde\omega_0| - |\tilde \omega|\right)
\right]   \nonumber\\
 &\ &\ \ \ +{ 1\over  2|\tilde \omega|}
     \left[\ln\left(1 +(|\tilde \omega| - |\tilde
\omega_0|)^2\right)             -\ln\left(1 +\tilde
\omega_0^2\right)\right]
      \bigg\}.
\label{Sscaled}
\end{eqnarray}
In Figure 6a, the excess noise $\tilde{S}-\tilde{S}^{\tilde{\omega}_0=0}$
is plotted against
$\tilde\omega/\tilde\omega_0$ for different values of $\tilde
\omega_0$.  As $\tilde \omega_0$ becomes large, the excess noise approaches
the step function in Eq. (\ref{SttsmallG}).  Recall that $\tilde\omega_0
= \tilde \omega/(4\pi|\Gamma|^2)$, so this limit is equal to the weak
coupling limit with $|\Gamma| \to 0$.  To see the strong coupling limit,
in Fig. 6b we plot the excess noise divided by $\tilde\omega_0^3$
(in order to fit in the same scale).  As
$\tilde\omega_0$ becomes small, this clearly has the cubic behavior in
Eq. (\ref{SttscatlG}).  Finally, the full noise, divided by $\tilde\omega_0$,
is plotted in Fig. 6c.  The cubic singularity at $\tilde\omega
= 2\tilde\omega_0$ decays too quickly to appear in the full noise.  However,
for $\tilde\omega_0 = 100$ and $\tilde\omega_0 = 10$, there is clearly a
``blip'' in the plot of the noise, which shows the ``smoothed out''
quasiparticle singularity.

For completeness, we will conclude this section by giving the
result for the noise $S(\omega;x_+,x_-)$ between incoming and
outgoing currents.  By comparing this with the perturbative
calculations of the cross-correlations, we will see to what extent
the duality symmetry holds.  In addition, once we have
$S(\omega;x_+,x_-)$,
$S(\omega;x_+,x_+)$ and $S(\omega;x_-,x_-)$, we can calculate the
noise in the Hall current and the tunneling current. The Hall
current is the total current running down the sample, given by
$I_H=j_L(x) + j_R(x)=\rho_R(x)-\rho_L(x)$ and the tunneling current
is the current that tunnels across the sample, which is given by
$I_t=\rho_R(x_+) -\rho_R(x_-)=\rho_L(x_-) -
\rho_L(x_+)$.

The expression for the $S(\omega;x_+,x_-)$ noise is
\begin{equation}
S(\omega;x_+,x_-)= \int_{-\infty}^\infty dt\ e^{i\omega t}
            \langle \{\rho_-(t, x_+), \rho_-(0, x_-)\} \rangle,
\label{Spmdef}
\end{equation}
where $x_- < 0$ and $x_+ > 0$.  Again, we can expand the $\rho_-
(x_+)$ and $\rho_-(x_-)$ in terms of the $A_\omega$'s and
$B_\omega$'s in the solution for $\psi$.  After evaluating the
expectation values and performing the integrals over $\omega_i$ and
$t$, we find
\begin{eqnarray}
S(\omega;x_+,x_-) =  \Bigg\{\frac{|\omega|}{2\pi}
       &-& 2 |\Gamma|^2
       \left[\tan^{-1}\left({|\omega|-\omega_0 \over4\pi |\Gamma|^2}\right)
           + \tan^{-1}\left({|\omega|+\omega_0
\over4\pi |\Gamma|^2}\right)\right]
 \nonumber  \\
       &+& i|\Gamma|^2 {\rm sign}(\omega)\bigg[
            2\ln\left(\omega_0^2 + (4\pi|\Gamma|^2)^2 \right)
    -   \ln\left((\omega + \omega_0)^2 + (4\pi |\Gamma|^2)^2 \right)
 \nonumber  \\
 &\ & \ \ \ \ \ \
     -   \ln\left((\omega - \omega_0)^2 + (4\pi |\Gamma|^2)^2 \right)\bigg]
  \bigg\} e^{i\omega(x_+ - x_-)}.
\label{Spm}
\end{eqnarray}
We can again use equation (\ref{rhorhoII}) to obtain the expression
for the cross-correlations of the currents in the original four
reservoirs.  We find, for example,
\begin{equation}
S_{31}(\omega) = {1\over 4}S(\omega; x_+, x_-)
          +{1\over 4}\frac{|\omega|}{2\pi} e^{i\omega (x_+ -x_-)}.
\end{equation}
and
\begin{equation}
S_{41}(\omega) = -{1\over 4}S(\omega; x_+, x_-)
          +{1\over 4}\frac{|\omega|}{2\pi} e^{i\omega (x_+ -x_-)}.
\end{equation}
The other cross-correlations, namely $S_{32}$  and
$S_{42}$, can be calculated similarly.  For small $|\Gamma|$, the
noise is
\begin{eqnarray}
S_{31}(\omega) = \bigg[{|\omega| \over 4\pi}
&-& {\pi\over4}|\Gamma|^2\left({\rm sign}(|\omega|+\omega_0) +
  {\rm sign}(|\omega|- \omega_0)\right) \nonumber \\
&-& i {1\over 2}|\Gamma|^2 \ln\left(\Big|{{\omega}^2 - \omega_0^2
     \over {\omega_0}^2}\Big|\right)\bigg]
                     e^{i\omega(x_+ - x_-)},
\label{Stosg}
\end{eqnarray}
and when $|\Gamma|$ is large, the noise becomes
\begin{eqnarray}
S_{41}(\omega) = \bigg\{{|\omega|\over 4\pi}
&+& i {1\over 32\pi^2}{1\over|\Gamma|^2} \omega^2{\rm sign}(\omega)
\nonumber \\
 &-&   {1\over 384\pi^3}{1\over|\Gamma|^4}
   \left[\left(|\omega|+\omega_0\right)^3
                +\left(|\omega|- \omega_0\right)^3\right]\bigg\}
 e^{i\omega(x_+ - x_-)}.
\end{eqnarray}
In the following section, we will compare these results with the
perturbative calculation.  We will find that for $g = 1/2$ they
agree, but they differ for $g=2$.  In Section V, we will also
discuss this apparent breakdown of the duality transformation.

\section{Discussion of the Duality Symmetry}
As we have seen in the previous sections, we expect this system to exhibit
a duality symmetry.  In this section, we will first describe this duality
symmetry more fully, and then compare the results from the perturbative and
scattering calculation to see how consistent they are with this symmetry.

For $g = 1/2$, the original picture of this system is a single quantum
Hall droplet with ``filling fraction'' $\nu = 1/2$.  Quasiparticles can
tunnel from one branch to the other, and they have charge $e^* = \nu e$,
tunneling amplitude $\Gamma_q$, and Josephson frequency $\omega_0 =
e^*V/\hbar$.  The Lagrangian describing this system can be written as
\begin{equation}
{\cal L} = {1\over 8\pi}\left[(\partial_t\phi)^2 - v^2(\partial_x \phi)^2
                 \right]
             -\Gamma_q e^{-i\omega_0 t} \delta(x) e^{i\sqrt g \phi(t,0)}
                      + H.c.,
\end{equation}
with $g = 1/2$ and $\phi = \phi_R(x) + \phi_L(-x)$.  If we use the
four probe geometry to study this system, then Eq. (\ref{rhonumdef}) gives the
relation between the densities in the four probes, $\rho_1$, $\rho_2$,
$\rho_3$, and $\rho_4$, and the densities of the leftmovers and rightmovers.
They are shown in Fig. 7a.

Once $\Gamma$ is increased (or $V$ is decreased), the droplet should split
into two.  Each of the two new droplets is still characterized by
filling fraction $\nu$.  However, now only electrons can tunnel across
the gap from one branch to the other.  For $g = 1/2$, the electron is made
up of two quasiparticles, so the tunneling operator for the electron should
be
\begin{equation}
\Gamma_e\left(e^{i\sqrt{1/2}\phi(t,0)}\right)^2
= \Gamma_e e^{i\sqrt{2}\phi(t,0)};
\end{equation}
the charge is $e$, and the Josephson frequency is $\tilde \omega_0 =
2\omega_0$.  Thus, when $\Gamma_q$ becomes large, this system can also be
described by the Lagrangian density
\begin{equation}
{\cal L} = {1\over 8\pi}\left[(\partial_t\phi)^2 - v^2(\partial_x \phi)^2
                 \right]
             -\Gamma_e e^{-i\tilde\omega_0 t} \delta(x)
                           e^{i\sqrt {\tilde g} \phi(t,0)}
                      + H.c.,
\end{equation}
where $\tilde g = 2$.  However, in this geometry with the two droplets, we
must be careful when we write the densities in the four probes in terms of
the left-moving and right-moving densities.  According to Fig. 7b, this
relation is given by
\begin{eqnarray}
\rho_1(t) &=& \rho_R(t,\tilde x) \qquad
       {\rm for} \quad \tilde x < 0 \nonumber\\
\rho_2(t) &=& \rho_L(t,\tilde x) \qquad
       {\rm for} \quad \tilde x > 0 \nonumber\\
\rho_3(t) &=& \rho_L(t,\tilde x) \qquad
       {\rm for} \quad \tilde x < 0 \nonumber\\
\rho_4(t) &=& \rho_R(t,\tilde x) \qquad
       {\rm for} \quad \tilde x > 0. \nonumber\\
\end{eqnarray}
With these identifications, $S_{31}(\omega)$ in the four probe geometry
equals $S_{LR}(\omega; \tilde x_-,\tilde x_-)$ in the two-droplet geometry,
and similarly, $S_{41}(\omega)$ is given by $S_{RR}(\omega; \tilde x_+,
\tilde x_-)$, where $\tilde x_- < 0$ and $\tilde x_+ > 0$.
Also, we see that the Hall current in the single droplet,
$\rho_R(t,x) - \rho_L(t,x)$ is dual to the tunneling current in the
two droplets, $\rho_R(t,\tilde x_-) - \rho_R(t,\tilde x_+)$, because both are
equivalent to $\rho_1(t) - \rho_4(t).$

We will first verify that the scattering and perturbative calculations
agree for $g = 1/2$.  We have already found that when the noise is
evaluated on only one side of the junction, then both the scattering
and  perturbative results agree.  If one probe is in an incoming channel
and the other probe is in an outgoing channel, then according to Appendix
A the perturbative result for the noise is
\begin{eqnarray}
S_{31}(\omega; x_+, x_-) = e^{i\omega(x_+-x_-)} \Bigg\{&\,&{|\omega|\over 4\pi}
   \nonumber\\
&-& {|\Gamma_q|^2\over 8}
\bigg[i4{\rm sign}(\omega)\ln\left(\left|{\omega^2-\omega_0^2\over \omega_0^2
           }\right|\right)
      + 2\pi \left[\left(1 + {\rm sign}\left(|\omega| - |\omega_0|\right)
                    \right)\right]\bigg]\Bigg\},
\end{eqnarray}
where we have set $g=\nu = 1/2$ in Eq. (\ref{SRRpert}).  On comparing this
with the expansion of the scattering calculation for small $\Gamma$
in Eq. (\ref{Stosg}), we find
that also in this case the scattering and perturbative results agree.

Next, to check the duality transformation, we must compare the scattering
calculation as $\Gamma \to \infty$ with the perturbative calculation at
$g=2$.  Again we found that if both probes are in the same branch then
the two calculations agree.  This is rather remarkable, because when
$g=2$ the system can be sensitive to short distance behavior, which
means that it could depend on the detailed structure of the junction
and on how it is regulated.  However, here we found that the
weak-coupling perturbative calculation and the strong-coupling limit
of the scattering calculation are the same, even though they treat
the junction very differently.  We conclude that, at least to the
order in perturbation theory that we have calculated, the noise
extracted from a single channel is not affected by the short-distance
properties of the impurity.

To complete the comparison, we need the results for the noise
between the incoming and outgoing channels.  Using Eq. (\ref{GtGoh}) to relate
the quasiparticle tunneling amplitude to the electron tunneling amplitude,
we find that the expansion for $\Gamma \to\infty$ of the scattering calculation
becomes
\begin{equation}
S_{41}(\omega; x_+, x_-) = \bigg\{{1\over 4\pi} |\omega|
           +{i\over2} |\Gamma_e|\omega^2 {\rm sign}(\omega)
          -{2\pi\over 3} |\Gamma_e|^2\left[\left(|\omega|+\omega_0\right)^3
                              +\left(|\omega|-\omega_0\right)^3\right]
                    \bigg\}e^{i\omega(x_+ - x_-)}.
\end{equation}
This must be compared with the perturbative calculation of
$S_{RR}(\omega; \tilde x_+, \tilde x_-)$.  To obtain this perturbative
result, we set $\nu = 1/2$, $g=2$ and replace $\omega_0$ by $2\omega_0$
in Eq. (\ref{SRRpert}).  Then the perturbative calculation of the noise
across the junction yields
\begin{equation}
S^{RR}(\omega; \tilde x_+, \tilde x_-) = \bigg\{{1\over 4\pi} |\omega|
    -|\Gamma_e|^2
     \left[{\pi\over 6}\left(\left(|\omega|+|2\omega_0|\right)^3
                             +\left(|\omega|-|2\omega_0|\right)^3\right)
                    +{2i\omega^2\over 3 \delta}\right]
                    \bigg\}e^{i\omega(x_+ - x_-)}.
\end{equation}
We first note that this expression for the noise contains a linear
divergence in the cutoff $\delta$.  Thus this perturbative calculation is
regulator dependent, which is not surprising  because the tunneling
operator at $g=2$ has dimension $2$ and should be an irrelevant operator.
In spite of this, both calculations do agree to order $\omega$ (which is all
that the derivations of the duality transformation in reference
\cite{Duality} would predict); it is only the higher order terms in $\omega$
that disagree.  This suggests that we are on the right track with the
perturbative calculation, but we just need to add in the appropriate counter
terms.

To see which counter term we should add, we begin by recalling that we used
the most relevant tunneling operator to describe the system.  However,
for $g=2$ the operators $(\rho_L)^2 + (\rho_R)^2$ and $\rho_L \rho_R$ are
just as relevant as the tunneling operator, so we must consider their
effects also.  In fact, $(\rho_R(0) - \rho_L(0))^2$ also encourages
tunneling because it tries to equalize the density of right movers and
left movers.  Another way to look at it is that we cannot have
quasiparticles tunneling between the droplets, but density fluctuations
on one side may affect the other side.

In Appendix B, we found that when the interaction
\begin{equation}
{\cal L}_{\rm int} = \gamma \left(\rho_R(t,0)-\rho_L(t,0)\right)^2
                     \delta(x)
\label{Lintdef}
\end{equation}
is included in the Lagrangian, it gives the following contribution to the
noise
\begin{equation}
S_{\rho\rho}^{RR}(x_1, x_2) = \theta(-x_1x_2)\bigg\{
               {i\gamma\over 8\pi^2}\omega^2{\rm sign}(\omega x_1)
              -{\gamma^2\over 16\pi^3} \left[|\omega|^3-
                        2i``\delta(0)'' \omega^2{\rm sign}(\omega x_1)\right]
               \bigg\} e^{i\omega(x_1-x_2)}.
\label{Srhorho}
\end{equation}
First, we note that the density-density coupling does not affect the noise
evaluated on only one side of the impurity (i.e. when $x_1 x_2 > 0$.)
According to equations (\ref{s11secIII}) to (\ref{s2secIII}),
(\ref{SttscatlG}), and (\ref{GtGoh}), this is necessary for the scattering and
the perturbative calculations to agree.  It is reasonable that the noise
evaluated on only one side of the junction should be less affected by the
counter terms and the regulator than the noise between probes on either side
of the junction, because even though in both cases all the measurements are
done far from the junction, in the second case the information must travel
from one side of the junction to the other.

Second, we note that when $x_1 x_2 < 0$, equation (\ref{Srhorho}) contains
the linear term in $\gamma$, which also appears in the scattering
calculation, but not in the original perturbative calculation.  We find that
the only density-density interaction that gives the same linear term as in the
scattering result for {\it all} of the cross-correlations is the one given in
equation (\ref{Lintdef}), with $\gamma = 1/(4\Gamma_q^2)$.  When we add the
density-density term with this choice for $\gamma$ to the original
perturbative calculation, we obtain
\begin{eqnarray}
S^{RR}_{\rm pert}(\omega; x_+, x_-) =
            \bigg\{{1\over 4\pi} |\omega|
           &+&{i\over2} |\Gamma_e|\omega^2 {\rm sign}(\omega)
          -{2\pi\over 3} |\Gamma_e|^2\left[\left(|\omega|+\omega_0\right)^3
                              +\left(|\omega|-\omega_0\right)^3\right]
    \nonumber \\
    &+&2i|\Gamma_e|^2\omega^2{\rm sign}(\omega)\left(\pi``\delta(0)''
        - {1\over 3\delta}\right)
                    \bigg\}e^{i\omega(x_+ - x_-)}.
\label{pernoise}
\end{eqnarray}
Thus, (except for the divergent part), this perturbative result agrees with
the scattering result.

To cancel the divergent part, we must regulate the
delta-function properly and adjust the counterterm accordingly.  Then the
two results will agree in the limit as $x_+$ and $x_- \to \pm \infty$.
Another approach, which may be more appropriate, is to ``smooth out''
the density-density interaction.  This is accomplished by replacing
the interaction in equation (\ref{Lintdef}) by the following expression
\begin{equation}
{\cal L}_{\rm int} = \gamma \left(\rho_R(t,0)-\rho_L(t,0)\right)^2
                     f_\epsilon(x),
\label{Lintf}
\end{equation}
where $f_\epsilon(x) \to \delta(x)$ as $\epsilon \to 0$.  This new
interaction does not change the finite part of Eq. (\ref{pernoise}),
and the function $f$ can be chosen so that the divergence cancels.
As a result, even though the duality symmetry is not exactly
obeyed for the cross-correlations, it is possible to add in counter terms to
bring the strong-coupling limit of one picture into agreement with the
weak-coupling limit of the dual picture.

To summarize, to the order in $\Gamma$ we have calculated, the action for
$g = 1/2$ is dual to the renormalized action for $g=2$, given by
\begin{equation}
{\cal L} = {1\over 8\pi}\left[(\partial_t\phi)^2 - v^2(\partial_x \phi)^2
                 \right]
             -\Gamma_e e^{-i2\omega_0 t} \delta(x)
                           e^{\sqrt {2} \phi(t,0)}
                      + H.c.
          +  4\pi^2\Gamma_e \delta(x)\left(\rho_R(t,0)-\rho_L(t,0)\right)^2;
\end{equation}
and if we only want to calculate the noise in one particular channel, then it
is not necessary to include the $\rho\rho$ interaction to obtain the dual
picture.  As explained above, this action can be interpreted as containing
two different terms that induce or encourage tunneling.  We can also use the
relation
\begin{equation}
\rho_R(0) -\rho_L(0) = {1\over 2 \sqrt{2} \pi} \partial_x\phi(0)
\end{equation}
to write the action as
\begin{equation}
{\cal L} = {1\over 8\pi}\left[(\partial_t\phi)^2-\tilde v^2(\partial_x \phi)^2
                 \right]
             -\Gamma_e e^{-i2\omega_0 t} \delta(x)
                           e^{\sqrt {\tilde 2} \phi(t,0)}
                      + H.c. ,
\end{equation}
where $\tilde v^2 = v^2 + 4\pi\delta(x) \Gamma_e^2$ is the ``renormalized''
velocity.  In this case, the velocity remains the same everywhere but
right at the junction.  If, instead, we use Eq. (\ref{Lintf}) for the density
interaction, then the velocity is renormalized in a region around the
junction.

\section{Conclusion}

In this paper we studied the four terminal tunneling noise spectrum
for chiral Luttinger liquids characterized by an exponent $g$.
Perturbative results are obtained for arbitrary $g$. Perturbative
calculations for quasiparticle tunneling reveal a singularity at the
quasiparticle Josephson frequency $\nu eV/\hbar$, while perturbative
calculations for electron tunneling only produce a singularity at the
electron Josephson frequency $eV/\hbar$. This appears to be
inconsistent with the duality picture that quasiparticle and electron
tunneling describe the same tunneling junction in two different limits.
To understand how the quasiparticle tunneling picture can smoothly
connect to the electron tunneling picture, we calculated the exact
noise spectrum for $g=1/2$ (or $g=2$ due to duality). We find that the
singularity at the quasiparticle Josephson frequency
$\frac{1}{2}eV/\hbar$ is smeared for finite tunneling and is not a true
singularity, while the singularity at the electron Josephson frequency
$eV/\hbar$ survives in the exact result. From the exact result we also
find that the noise spectrum of the current correlations on a single
branch (auto-correlations) satisfies the duality relation, while
current correlations between distinct branches (cross-correlations) do
not satisfy the naive duality relation.

\

\begin{center}
{\bf ACKNOWLEDGEMENTS}
\end{center}

We would like to thank Akira Furusaki for helpful discussions.
This work is supported by NSF grants  DMR-9400334 (CCC) and DMR-9411574
(XGW).
XGW acknowledges the support from the A.P. Sloan Foundation.
D.~F.~is currently a
Bunting Fellow sponsored by the Office of Naval Research
and is also supported in part by
funds provided by the U. S. Department of Energy (D.O.E.) under cooperative
agreement \#DE-AC02-76ER03069 and
by National Science Foundation grant PHY9218167.

\appendix
\section{perturbative calculation}

In order to obtain the noise spectrum of density-density
correlations on given leads, we start by writing the correlations
between density operators as follows:
\begin{equation}
\langle \rho_{a}(t,x_1) \rho_{b}(0,x_2) \rangle\ ,
\end{equation}
where $a,b$ take the values $+1$ for $R$ moving branches and $-1$
for $L$ moving ones. Such compressed notation makes it simpler to
identify incoming and outgoing branches in a unified way for both
left and right movers: $\rho_a(t,x_1)$, for example, is the density
in an incoming or outgoing branch if $a x_1<0$ or $a x_1>0$,
respectively.

The densities are related to the fields $\phi_{R,L}$ through
$\rho_{R,L}=\frac{\sqrt{\nu}}{2\pi}\partial_x \phi_{R,L}$, so that
we can write
\begin{equation}
\langle
\rho_{a}(t,x_1) \rho_{b}(t',x_2) \rangle =
\frac{\nu}{(2\pi)^2}\ \partial_{x_1}\partial_{x_2}\langle
\phi_{a}(t,x_1)\phi_{b}(t',x_2) \rangle\ ,
\end{equation}
where it is convenient to use
\begin{equation}
\langle \phi_{a}(t,x_1)\phi_{b}(t',x_2)
\rangle=
\frac{d\ }{d\lambda_1}\frac{d\ }{d\lambda_2}
\langle
e^{i\lambda_1\phi_a(t,x_1)}\ e^{-i\lambda_2\phi_b(t',x_2)}
\rangle{\large |}_{\lambda_1,\lambda_2=0}\ .\label{phiphi}
\end{equation}
The last correlation function is easy to calculate perturbatively
using
\begin{equation}
\langle T_c(e^{i\lambda_1\phi_a(t,x_1)}\ e^{-
i\lambda_2\phi_b(t',x_2)}) \rangle=
\langle 0|\ T_c(S(-\infty,-\infty)\ e^{i\lambda_1\phi_a(t,x_1)}\
e^{-i\lambda_2\phi_b(t',x_2)}) \ |0\rangle\ ,
\end{equation}
where $|0\rangle$ is the unperturbed ground state, and $T_c$ is the
ordering along the Keldysh contour (Fig. 8). The scattering
operator $S(-\infty,-\infty)$ takes the initial state, evolves it
from $t=-\infty$ to $t=\infty$ and back to $t=-\infty$. The use of
the Keldysh contour is necessary in the treatment of non-
equilibrium problems, such as the one we have in hand. A more
detailed description of the method in the context treated here can
be found in Ref. \cite{CFW}.

In order to proceed we expand $S(-\infty,-\infty)$ to second order
in perturbation theory. In terms of the Coulomb gas of
Ref. \cite{CFW}, we have an insertion of two charges of opposite
sign:
\begin{eqnarray}
\langle &T_c&(e^{i\lambda_1\phi_a(t,x_1)}\ e^{-
i\lambda_2\phi_b(t',x_2)}) \rangle_{|\Gamma|^2}=\\
&\ &(i\Gamma)(i\Gamma^*)
\oint_c dt_+ \oint_c dt_- \ e^{i\omega_0t_+}\ e^{-i\omega_0t_-}
\langle 0|\ T_c(\ e^{iq\phi(t_+,0)}\ e^{-iq\phi(t_-,0)}\
e^{i\lambda_1\phi_a(t,x_1)}\ e^{-i\lambda_2\phi_b(t',x_2)}) \
|0\rangle\ ,\nonumber
\end{eqnarray}
where $q=\sqrt{g}$, and $\phi$ without subscript stands for the sum
$\phi_R+\phi_L$.  The expression above is simplified using
\begin{equation}
\langle 0|T_c (\prod_j e^{i q_j\phi(t_j,x_j)})|0 \rangle=
e^{-\sum_{i> j}q_iq_j
\langle
0|T_c(\phi(t_i,x_i)\phi(t_j,x_j))|0\rangle}\ .
\end{equation}
Substituting it into Eq. (\ref{phiphi}) we obtain
\begin{eqnarray}
\langle T_c(\phi_{a}(t,x_1)\phi_{b}(t',x_2))
\rangle_{|\Gamma|^2} &=&|\Gamma|^2
\oint dt_+ \oint dt_- e^{q^2\langle 0|T_c(\phi(t_+,0)\phi(t_-
,0))|0\rangle} e^{i\omega_0(t_+-t_-)}\nonumber\\
&\ &\ \times \{q^2
\left[\langle 0|T_c(\phi(t_+,0)\phi_a(t,x_1))|0\rangle -
\langle 0|T_c(\phi(t_-,0)\phi_a(t,x_1))|0\rangle\right]\nonumber\\
&\ &\ \ \ \ \ \ \times\left[\langle
0|T_c(\phi(t_+,0)\phi_b(t',x_2))|0\rangle -\langle 0|T_c(\phi(t_-
,0)\phi_b(t',x_2))|0\rangle\right]\nonumber\\ &\ &\ \ \ \ \ \ +
\langle 0|T_c(\phi_a(t,x_1)\phi_b(t',x_2))|0\rangle\}\
.\label{correl}
\end{eqnarray}
The last term in the expression above, the one proportional to
$\langle 0|T_c(\phi_a(t,x_1)\phi_b(t',x_2))|0\rangle\}$, vanishes.
The reason why this happens is very simple: the factor in front of
it is the term of order $|\Gamma|^2$ in the expansion of $Z=\langle
0|S(-\infty,-\infty)|0\rangle$; since $Z\equiv 1$, the correction
at any order in $\Gamma$ must vanish.

In order to carry out the calculations, we introduce notation
that keeps track of the position of the two inserted charges along
the contour, {\it i.e.}, whether they are in the forward (or top)
branch, or in the return (or bottom) branch (see Fig. 8). The
position of the charges is important for the computation of the
contour-ordered correlation function, given by $\langle
0|T_c(\phi_{R,L}(t_1,x_1)\phi_{R,L}(t_2,x_2))|0\rangle$
\[ =\left\{ \begin{array}{ll}
-\ln (\delta +i\ {\rm sign}(t_1-t_2)[(t_1-t_2)\mp(x_1-x_2)],&\mbox{
both $t_1$ and $t_2$ in the top branch}\\ -\ln (\delta -i\ {\rm
sign}(t_1-t_2)[(t_1-t_2)\mp(x_1-x_2)],&\mbox{ both $t_1$  and $t_2$
in the bottom branch}\\
-\ln (\delta -i[(t_1-t_2)\mp(x_1-x_2)]),&\mbox{ $t_1$  in the top
and $t_2$ in the bottom branch}\\
-\ln (\delta +i[(t_1-t_2)\mp(x_1-x_2)]),&\mbox{ $t_1$  in the
bottom and $t_2$ in the top branch.}
\end{array}
\right. \]
\\
The compact notation consists of giving indices to the times which contain
the information about which branch of the Keldysh contour they are on,
so that $t^\mu$ is on the top branch for $\mu=+1$, and on the bottom for
$\mu=-1$. In this way, we can compress the correlations to a
compact form:
\begin{eqnarray}
G_{\mu\nu}^{ab}(t_1,x_1;t_2,x_2)&=&G_{\mu\nu}^{ab}(t_1-t_2,x_1-
x_2)= \langle
0|T_c(\phi_a(t^\mu_1,x_1)\phi_b(t^\nu_2,x_2))|0\rangle\nonumber\\
&=&- \delta_{a,b}\ \ln (\delta +i\ K_{\mu\nu}(t_1-t_2)[(t_1-t_2)-
a(x_1-x_2)])\ ,\label{G}
\end{eqnarray}
where
\begin{equation}
K_{\mu\nu}(t)=\theta(\mu\nu)\  {\rm sign}(\nu t)+
\theta(-\mu\nu)\  {\rm sign}(\nu)\ .\label{K}
\end{equation}
Again, we have used $a,b=\pm 1$ for $R$ and $L$ fields,
respectively. The correlation in Eq. (\ref{correl}) can be written,
using this compressed notation, as
\begin{eqnarray}
\langle T_c(\phi_{a}(t,x_1)\phi_{b}(t',x_2))
\rangle_{|\Gamma|^2}
&=&|\Gamma|^2 q^2\sum_{\mu\nu}{\rm sign}(\mu\nu)\int_{-
\infty}^\infty dt_+\int_{-\infty}^\infty dt_-  \ e^{i\omega_0(t_+-
t_-)}P_{\mu\nu}(t_+-t_-)\nonumber\\
&\ &\ \ \ \times
[G_{+\mu}^{aa}(t-t_+,x_1)-G_{+\nu}^{aa}(t-t_-,x_1)]\nonumber\\ &\
&\ \ \ \times
[G_{+\mu}^{bb}(t'-t_+,x_2)-G_{+\nu}^{bb}(t'-t_-,x_2)]\
,\label{corG}
\end{eqnarray}
where $P_{\mu\nu}(t_+-t_-)=e^{q^2[G_{\mu\nu}^{++}(t_+-t_-,0)
+G_{\mu\nu}^{--}(t_+-t_-,0)]}$, or explicitly:
\begin{equation}
P_{\pm\pm}(t)=\frac{1}{(\delta \pm i|t|)^{2g}}\ ,\
P_{\pm\mp}(t)=\frac{1}{(\delta \mp it)^{2g}}\ .
\end{equation}
The factor ${\rm sign}(\mu\nu)$ simply keeps track of the sign
coming from the integration of the times $t_\pm$ along the contour.
Notice that the times $t$ and $t'$ are taken to be on the top
branch.

Now, let
\begin{eqnarray}
F_{ab}(\omega;x_1,x_2)&=&\int_{-\infty}^\infty dt\
e^{i\omega t}\ \langle T_c(\rho_{a}(t,x_1)\rho_{b}(0,x_2))
\rangle_{|\Gamma|^2}\nonumber\\
&=&\frac{\nu}{(2\pi)^2}\ \partial_{x_1}\partial_{x_2}
\int_{-\infty}^\infty dt\
e^{i\omega t}\ \langle T_c(\phi_{a}(t,x_1)\phi_{b}(0,x_2))
\rangle_{|\Gamma|^2}\ ,
\end{eqnarray}
which can be easily shown, using Eq. (\ref{corG}), to yield
\begin{eqnarray}
F_{ab}(\omega;x_1,x_2)=|\Gamma|^2 \frac{\nu q^2}{(2\pi)^2}
\sum_{\mu\nu}{\rm sign}(\mu\nu)
\big[ &\ &{\tilde P_{\mu\nu}(\omega_0)}
\left( g_{+\mu}^{aa}(\omega,x_1)g_{+\mu}^{bb}(-\omega,x_2)+
g_{+\nu}^{aa}(\omega,x_1)g_{+\nu}^{bb}(-
\omega,x_2)\right)\nonumber\\ -&\ &{\tilde P_{\mu\nu}(\omega_0-
\omega)}\
g_{+\mu}^{bb}(-\omega,x_2)g_{+\nu}^{aa}(\omega,x_1)\nonumber\\ -&\
&{\tilde P_{\mu\nu}(\omega_0+\omega)}\
g_{+\mu}^{aa}(\omega,x_1)g_{+\nu}^{bb}(-\omega,x_2)\ \big]
\ .\label{F}
\end{eqnarray}
In this equation, $g$ is given by
$g_{+\mu}^{aa}(\omega,x)=\partial_{x}{\tilde
G}_{+\mu}^{aa}(\omega,x)$ and  can be obtained from Eqs.
(\ref{G}) and (\ref{K}):
\[ g_{\mu\nu}^{ab}(\omega,x)=\delta_{a,b}\times \left\{
\begin{array}{ll} \pi ia\ e^{i\omega ax}\ \left({\rm
sign}(\omega)+{\rm sign}(ax)\right)&\mbox{, $\mu=+1,\nu=+1$}\\ \pi
ia\ e^{i\omega ax}\ \left({\rm sign}(\omega)-{\rm
sign}(ax)\right)&\mbox{, $\mu=-1,\nu=-1$}\\ -2\pi ia\ e^{i\omega
ax}\ \theta(-\omega)&\mbox{, $\mu=+1,\nu=-1$}\\ 2\pi ia\ e^{i\omega
ax}\ \theta(\omega)&\mbox{, $\mu=-1,\nu=+1$}\\ \end{array}
\right. \]
\\
The spectrum to second order can be obtained from
$F_{ab}(\omega,x_1,x_2)$ as follows:
\begin{eqnarray}
S^{(2)}_{ab}(\omega;x_1,x_2)&=&S^{(2)}_{ba}(-
\omega;x_2,x_1)=\int_{-\infty}^\infty dt\ e^{i\omega t} \langle
\{\rho_a(t,x_1),\rho_b(0,x_2)\}\rangle_{|\Gamma|^2}\nonumber\\
&=&F_{ab}(\omega;x_1,x_2)+F^*_{ab}(-\omega;x_1,x_2)\ .\label{S}
\end{eqnarray}

The only ingredients remaining to be calculated are the
$P(\omega)'s$, which are given by:
\begin{eqnarray}
{\tilde P_{++}(\omega)}&=&t(-\omega)\ =\int_{-\infty}^\infty dp
\frac{e^{i\omega p}}{(\delta + i|p|)^{2g}}\nonumber\\ {\tilde P_{--
}(\omega)}&=&b(-\omega)\ =\int_{-\infty}^\infty dp \frac{e^{i\omega
p}}{(\delta - i|p|)^{2g}}\\ {\tilde P_{\pm\mp}(\omega)}&=&c_\pm(-
\omega)=\int_{-\infty}^\infty dp \frac{e^{i\omega p}}{(\delta \mp
ip)^{2g}}\ .\nonumber \end{eqnarray}
The $t,b,c_\pm$ are the same as  in Ref. \cite{CFW}. One
can easily check that $t(\omega)+b(\omega)=c_+(\omega)+c_-
(\omega)$, and that the $c_\pm$ are given by
\begin{equation}
c_{\pm}(\omega)= \int_{-\infty}^\infty dp \frac{e^{-i\omega
p}}{(\delta \mp ip)^{2g}}
=\frac{2\pi}{\Gamma(2g)}|\omega|^{2g-1}e^{-|\omega|\delta} \ \theta
(\pm\omega) \ .
\end{equation}

Now, we have the tools we need in order to obtain all correlations.
In particular, correlations within the same branch and taken at the
same point, {\it i.e.}, $a=b$ and $x_1=x_2$, can be shown to yield:
\begin{equation}
S^{(2)}_{aa}(\omega;x_1,x_1)=\frac{4\pi\nu g}{\Gamma(2g)}\
|\Gamma|^2\ \theta(ax_1)\ \Big||\omega|-|\omega_0|\Big|^{2g-1} \
\theta(|\omega_0|-|\omega|)\ .
\end{equation}
The zero order term in $\Gamma$ is trivially obtained from the
unperturbed density-density correlation functions:
\begin{eqnarray}
S^{(0)}_{ab}(\omega;x_1,x_2)&=&S^{(0)}_{ba}(-\omega;x_2,x_1)=
\int_{-\infty}^\infty dt\ e^{i\omega t}
\langle 0|\{\rho_a(t,x_1),\rho_b(0,x_2)\}|0\rangle\nonumber\\
&=&\frac{\nu}{2\pi}|\omega|\ \delta_{a,b}\
e^{i\omega a(x_1-x_2)}\ ,\label{0thorder}
\end{eqnarray}
so that, in particular,
$S^{(0)}_{aa}(\omega;x_1,x_1)=\frac{\nu}{2\pi}|\omega|$.

Combining the zeroth and second order results, we obtain the
results used in section III for the noise in incoming ($ax_1<0$)
and outgoing ($ax_1>0$) branches, namely:
\[ S(\omega)=\left\{ \begin{array}{ll}
\ \frac{\nu}{2\pi}|\omega|,&\mbox{incoming branches}\\
\frac{\nu}{2\pi}|\omega| + \frac{4\pi\nu g}{\Gamma(2g)}\
|\Gamma|^2\ \ \Big||\omega| -|\omega_0|\Big|^{2g-1}\
\theta(|\omega_0|-|\omega|),&\mbox{outgoing branches}\\
\end{array}
\right. \]
\\
It is straightforward to show that the noise in the incoming branch remains
equal to ${\nu\over 2\pi}|\omega|$ to all orders in perturbation theory.

Next, we will obtain correlations between densities of an incoming
and an outgoing branch (the cross-correlations). Without loss of
generatity, we will focus on the correlations between two $R$
branches ($a=1$), one outgoing ($x_1>0$), and another incoming
($x_2<0$). The results for other combinations of branches are
trivially obtained from the case we consider.  We have, again, all
the tools at hand, namely Eqs. (\ref{F}) and (\ref{S}), as well as
our expressions for
$g_{\mu\nu}^{ab}(\omega,x)$ and ${\tilde P_{\mu\nu}(\omega)}$. We
find
\begin{eqnarray}
S^{(2)}_{RR}(\omega;x_1>0,x_2<0)&=&
\int_{-\infty}^\infty dt\ e^{i\omega t}
\langle
\{\rho_R(t,x_1>0),\rho_R(0,x_2<0)\}\rangle_{|\Gamma|^2}\nonumber\\
&=&
e^{i\omega(x_1-x_2)}\ \frac{|\Gamma|^2g\nu}{2}
\Bigg\{ {\rm sign}(\omega)H_g(\omega)\nonumber \\
&-&\frac{2\pi}{\Gamma(2g)}
\left[ (|\omega|+|\omega_0|)^{2g-1}+\Big||\omega|-
|\omega_0|\Big|^{2g-1} {\rm sign}(|\omega|-
|\omega_0|)\right]\Bigg\}\ ,
\label{SRRpert}
\end{eqnarray}
where the function $H_g(\omega)$ is defined as
\begin{eqnarray}
H_g(\omega)&=&2[t(\omega_0)-b(\omega_0)]
-[t(\omega_0-\omega)-b(\omega_0-\omega)]
-[t(\omega_0+\omega)-b(\omega_0+\omega)]\nonumber \\
&=&8 \int_0^\infty dt\ \cos(\omega_0 t)\sin^2(\omega t/2)
\left[\frac{1}{(\delta+it)^{2g}}-\frac{1}{(\delta-it)^{2g}}\right]\ .
\end{eqnarray}
One can show particularly that
$H_{1/2}(\omega)=-4i\ln(|\frac{\omega^2-\omega_0^2}{\omega_0^2}|)$,
$H_{1}(\omega)=0$, and
$H_{2}(\omega)=-\frac{4i\omega^2}{3\delta}\rightarrow\infty$ as
$\delta\rightarrow 0$.

The zero order contribution to the cross-correlations
is read directly from Eq.(\ref{0thorder}):
$S^{(0)}_{RR}(\omega;x_1>0,x_2<0)=\frac{\nu}{2\pi}|\omega|\
e^{i\omega (x_1-x_2)}\ $.

\section{perturbative calculation for the density-density coupling}

Here we consider the neutral coupling $L_{\rm int}=\gamma(\rho_R(t,0)
-\rho_L(t,0))^2$, and show that it contributes to the correlations
between incoming and outgoing branches, although it does not
contribute to correlations between two incoming or two outgoing
ones.
The calculations are simpler than the ones in Appendix A. We will
demonstrate the point by calculating the correlation $\langle
T_c(\rho_{R}(t,x_1) \rho_{R}(0,x_2)) \rangle$ to first and second
order in $\gamma$. Other correlations can be calculated in a very
similar way.

As in Appendix A, contour integrals are simplified by keeping track
of insertions in the top and bottom branches with indices
$\mu,\nu=\pm 1$. It is useful to define
\begin{eqnarray}
h_{\mu\nu}^{ab}(t_1,x_1;t_2,x_2)&=&h_{\mu\nu}^{ab}(t_1-t_2,x_1-
x_2)= \langle
0|T_c(\rho_a(t^\mu_1,x_1)\rho_b(t^\nu_2,x_2))|0\rangle\nonumber\\
&=&
\frac{\nu}{(2\pi)^2}\partial_{x_1}\partial_{x_2}
\langle 0|T_c(\phi_a(t^\mu_1,x_1)\phi_b(t^\nu_2,x_2))|0\rangle\\
&=&
\frac{\nu}{(2\pi)^2}\partial_{x_1}\partial_{x_2}
G_{\mu\nu}^{ab}(t_1-t_2,x_1-x_2)\ ,\nonumber\\
\end{eqnarray}
where $a$ and $b$, as in Appendix A, take the values $+1$ for $R$ moving
branches and $-1$ for $L$ moving ones. It follows from the
calculations of Appendix A that
${\tilde h}_{\mu\nu}^{ab}(\omega,x)=
- \frac{\nu}{(2\pi)^2}\partial^2_{x}{\tilde G}_{\mu\nu}^{ab}(\omega,x)
= - \frac{\nu}{(2\pi)^2}\partial_{x}g_{\mu\nu}^{ab}(\omega,x)$,
which gives
\[ {\tilde
h}_{\mu\nu}^{ab}(\omega,x)=\frac{\nu}{(2\pi)^2}\delta_{a,b}\times
\left\{ \begin{array}{ll} 2\pi |\omega|\ \theta(a\omega x)\
e^{i\omega ax} \ -2\pi
i\delta(x),&\mbox{$\mu=+1,\nu=+1$}\\ 2\pi |\omega|\ \theta(-
a\omega x)\ e^{i\omega ax} +2\pi i\delta(x),&\mbox{
$\mu=-1,\nu=-1$}\\ 2\pi |\omega|\ \theta(-\omega)\ e^{i\omega
ax},&\mbox{$\mu=+1,\nu=-1$}\\ 2\pi |\omega|\ \theta(\omega)\
e^{i\omega ax},&\mbox{
$\mu=-1,\nu=+1$}\\
\end{array}
\right. \]
\\
The perturbative results can be easily written in terms of these
${\tilde h}$'s.

Notice that the only term in the interaction
$\rho_-^2=\rho_R^2-2\rho_R\rho_L+\rho_L^2$ that contributes to
$\langle T_c(\rho_{R}(t,x_1) \rho_{R}(0,x_2)) \rangle$ to order
$\gamma$ is the $\rho_R^2$ term. The first order in $\gamma$
correction to the correlation function can be written as
\begin{eqnarray}
\langle &T_c&(\rho_{R}(t,x_1) \rho_{R}(0,x_2))
\rangle_{\gamma}\nonumber\\ &=&
i\gamma \oint_c dt_1\ \
\langle 0|\
T_c(\rho_{R}(t,x_1) \rho_{R}(0,x_2)
\rho_{R}(t_1,0)\rho_{R}(t_1,0))
\ |0\rangle\nonumber\\
&=&
2i\gamma\oint_c dt_1\ \
\langle 0|\ T_c(\rho_{R}(t,x_1)\rho_{R}(t_1,0))\ |0\rangle
\times
\langle 0|\ T_c(\rho_{R}(0,x_2)\rho_{R}(t_1,0))\ |0\rangle\ .
\label{r-r,o1}
\end{eqnarray}
The Fourier transform $F^{(1)}_{RR}(\omega;x_1,x_2)$ of the
expression in Eq.(\ref{r-r,o1}) is simply
\begin{equation}
F^{(1)}_{RR}(\omega;x_1,x_2)=2i\gamma\sum_{\mu}{\rm sgn }(\mu)\
{\tilde h}_{+\mu}^{++}(\omega,x_1){\tilde h}_{+\mu}^{++}(-
\omega,x_2)\ ,
\end{equation}
and thus to first order in $\gamma$ the cross-correlation spectrum
is given by
\begin{eqnarray}
S^{(1)}_{RR}(\omega;x_1,x_2)&=&S^{(1)}_{RR}(-\omega;x_2,x_1)
=\int_{-\infty}^\infty dt\ e^{i\omega t}
\langle \{\rho_R(t,x_1),\rho_R(0,x_2)\}\rangle_\gamma\nonumber\\
&=&F^{(1)}_{RR}(\omega;x_1,x_2)+{F^{(1)}_{RR}}^*(-
\omega;x_1,x_2)\nonumber\\ &=&\frac{i\gamma\nu^2}{2\pi^2}\ \theta(-
x_1x_2)\
\omega^2\ {\rm sign}(\omega x_1)\ e^{i\omega(x_1-x_2)}\ .
\end{eqnarray}

Turning now to second order in the perturbation expansion, both the
$\rho_R^2$ and the $\rho_R\rho_L$ terms in the interaction
$\rho_-^2=\rho_R^2-2\rho_R\rho_L+\rho_L^2$ can contribute to the
order $\gamma^2$ correction to $\langle T_c(\rho_{R}(t,x_1)
\rho_{R}(0,x_2)) \rangle$. Consider the $\gamma\rho_R^2$ coupling,
so that to second order we have
\begin{eqnarray}
\langle &T_c&(\rho_{R}(t,x_1) \rho_{R}(0,x_2))
\rangle_{(\gamma\rho_R^2)^2}\nonumber\\ &=&
\frac{(i\gamma)^2}{2!}\oint_c dt_1 \oint_c dt_2
 \langle 0|\
T_c(\rho_{R}(t,x_1) \rho_{R}(0,x_2)
\rho_{R}(t_1,0)\rho_{R}(t_1,0)\rho_{R}(t_2,0)\rho_{R}(t_2,0)) \
|0\rangle\nonumber\\
&=&8\frac{(i\gamma)^2}{2!}
\oint_c dt_1 \oint_c dt_2\ \
\langle 0|\ T_c(\rho_{R}(t,x_1)\rho_{R}(t_1,0))\
|0\rangle\nonumber\\ &\ &\ \ \ \ \ \ \ \ \ \ \ \ \ \ \ \ \ \ \ \ \
\ \ \ \ \ \ \times \langle 0|\ T_c(\rho_{R}(0,x_2)\rho_{R}(t_2,0))\
|0\rangle\nonumber\\ &\ &\ \ \ \ \ \ \ \ \ \ \ \ \ \ \ \ \ \ \ \ \
\ \ \ \ \ \ \ \ \times \langle 0|\
T_c(\rho_{R}(t_1,0)\rho_{R}(t_2,0))\ |0\rangle\ .\label{r-r,o2}
\end{eqnarray}
The effect of the $\gamma 2\rho_R\rho_L$ coupling can be calculated
likewise. The Fourier transform of these two contributions combined
give
\begin{equation}
F^{(2)}_{RR}(\omega;x_1,x_2)=-4\gamma^2\sum_{\mu\nu}{\rm
sign}(\mu\nu) {\tilde h}_{+\mu}^{++}(\omega,x_1){\tilde
h}_{+\nu}^{++}(-\omega,x_2) \left[ {\tilde
h}_{\mu\nu}^{++}(\omega,0)+{\tilde h}_{\mu\nu}^{--}(\omega,0)
\right]\ ,
\end{equation}
so that the cross-correlation spectrum to second order is
\begin{eqnarray}
S^{(2)}_{RR}(\omega;x_1,x_2)&=&S^{(2)}_{RR}(-\omega;x_2,x_1)
=\int_{-\infty}^\infty dt\ e^{i\omega t}
\langle
\{\rho_R(t,x_1),\rho_R(0,x_2)\}\rangle_{\gamma^2}\nonumber\\
&=&F^{(2)}_{RR}(\omega;x_1,x_2)+{F^{(2)}_{RR}}^*(-
\omega;x_1,x_2)\nonumber\\ &=&
-\frac{\gamma^2\nu^3}{2\pi^3}\ \theta(-x_1x_2)\ e^{i\omega(x_1-
x_2)} \left[|\omega|^3 - 2i\ ``\delta(0)''\ \omega^2\ {\rm
sign}(\omega x_1) \right]\ ,
\end{eqnarray}
where ``$\delta(0)$'' is a regulation dependent divergent term.

Notice that, both to first and second order in $\gamma$, the
correlations on the same side of the junction, {\it i.e.},
$x_1x_2>0$, do not feel the density-density coupling, whereas
correlations across the junction ($x_1x_2<0$) do feel the coupling.
More generally, when one considers all possible correlations
involving $R$ and $L$ branches, only those which contain an
incoming and an outgoing branch will have a non-zero correction due
to the density-density coupling.
Correlations between two incoming or two outgoing branches will be
zero.


\section{Scattering Calculation}
In this appendix, we evaluate the expectation values and integrals
used for calculating the noise in Section IV.  The methods of
calculation are very similar to those in \cite{Buttiker}

First, we will evaluate the noise in the incoming reservoir, which
is  given by
\begin{equation}
S(\omega; x_-, x_-) = \int_{-\infty}^\infty dt
      \left(e^{i\omega t} + e^{-i\omega t}\right)
            \langle \rho_-(t, x_-) \rho_-(0, x_-) \rangle.
\label{Smmdefc}
\end{equation}
The expectation value we must calculate is given by
\begin{equation}
\langle \rho_-(t, x_-) \rho_-(0, x_-) \rangle
  = \langle \psi^\dagger(t,x_-)\psi(t, x_-)\psi^\dagger(0,x_-
)\psi(0,x_-)      \rangle,
\label{ImImdefc}
\end{equation}
with $x_- < 0$.  Using the solutions in Eqs. (\ref{psixminus}) and
(\ref{psidxminus}) for $\psi$ and $\psi^\dagger$, we find
\begin{equation}
\langle \rho_-(t, x_-) \rho_-(0, x_-) \rangle
  = \sum_{\omega_1,\omega_2,\omega_3,\omega_4} e^{-
i(\omega_1+\omega_2)t}      \langle \Phi |
     A_{-\omega_1}^\dagger A_{\omega_2} A_{-\omega_3}^\dagger
A_{\omega_4}      | \Phi \rangle
      e^{i(\omega_1 + \omega_2 + \omega_3 + \omega_4) x_-}.
\label{ImImcalcc}
\end{equation}
The connected part of
$\langle A_{-\omega_1}^\dagger A_{\omega_2} A_{-\omega_3}^\dagger
A_{\omega_4} \rangle$ has $A^\dagger_{-\omega_1}$ paired with
$A_{\omega_4}$  and $A_{\omega_2}$ paired with $A^\dagger_{-
\omega_3}$ and is given by
\begin{equation}
 \langle \Phi |
     A_{-\omega_1}^\dagger A_{\omega_2} A_{-\omega_3}^\dagger
A_{\omega_4}  | \Phi \rangle_{\rm con}
   = \langle \Phi | A_{-\omega_1}^\dagger A_{\omega_4} | \Phi
\rangle      \langle \Phi | A_{\omega_2} A_{-\omega_3}^\dagger  |
\Phi \rangle. \label{AAAAcalc}
\end{equation}
Evaluating these correlations using equations (\ref{Acom}) and
(\ref{AdAexp}), we find that the current-current correlation
reduces to
\begin{equation}
\langle \rho_-(t, x_-) \rho_-(0, x_-) \rangle
  = \sum_{\omega_1,\omega_2} e^{-i(\omega_1+\omega_2)t}
     n_{-\omega_1}(1- n_{\omega_2}).
\label{ImImn}
\end{equation}
Substituting this expression back into equation (\ref{Sjjdef}) for
the noise,  and performing the integrals over $t$ and $\omega_1$,
we obtain
\begin{equation}
 S(\omega; x_-, x_-) (\omega) = \int_{-\infty}^\infty
\frac{d\omega_2}{2\pi}\  n_{\omega_2-\omega}(1- n_{\omega_2})    +
\int_{-\infty}^\infty \frac{d\omega_2}{2\pi}
     n_{\omega_2+\omega}(1- n_{\omega_2}).
\label{Smmn}
\end{equation}
At zero temperature, the integrands are given by
\begin{equation}
n_{\omega_2\mp\omega}(1- n_{\omega_2}) = \cases{
          1 & for $\pm \omega > 0$ and
                  $\omega_0 \le \omega_2 \le \omega_0 \pm \omega$
\cr           0 & otherwise. \cr}
\label{nnlimits}
\end{equation}
Performing the integral, we obtain the desired result:
\begin{equation}
S(\omega; x_-, x_-)(\omega) = \frac{1}{2\pi}|\omega|.
\label{Smmansc}
\end{equation}

Next, we will calculate the noise in the outgoing current, which is
given by
\begin{equation}
S(\omega; x_+, x_+) = \int_{-\infty}^\infty dt
      \left(e^{i\omega t} + e^{-i\omega t}\right)
            \langle \rho_-(t, x_+) \rho_-(0, x_+) \rangle,
\label{Sppdefc}
\end{equation}
where $x_+ > 0$.
This time we must evaluate the expectation value
\begin{equation}
\langle \rho_-(t, x_+) \rho_-(0, x_+) \rangle
  = \langle \psi^\dagger(t,x_+)\psi(t,
x_+)\psi^\dagger(0,x_+)\psi(0,x_+)      \rangle,
\label{IpIpdefc}
\end{equation}
According to equations (\ref{psixplus}) and (\ref{psidxplus}), this
is equal to
\begin{equation}
\langle \rho_-(t, x_+) \rho_-(0, x_+) \rangle
  = \sum_{\omega_1,\omega_2,\omega_3,\omega_4} e^{-
i(\omega_1+\omega_2)t}      \langle \Phi |
     B_{-\omega_1}^\dagger B_{\omega_2} B_{-\omega_3}^\dagger
B_{\omega_4}      | \Phi \rangle
      e^{i(\omega_1 + \omega_2 + \omega_3 + \omega_4) x_+}.
\label{IpIpcalcc}
\end{equation}
Because the scattering states are defined in terms of the operator
$A_\omega$, we will use equation (\ref{Bdef}) to rewrite all the
$B$'s in terms of the $A$'s, with the result
\begin{equation}
     \langle
     B_{-\omega_1}^\dagger B_{\omega_2} B_{-\omega_3}^\dagger
B_{\omega_4}      \rangle  = s_o + s_t.
\label{BBBBSoSt}
\end{equation}
In this equation, $s_o$ describes events where at both time $0$
and at time $t$ one
particle is destroyed and another is created. The second term,
$s_t$, describes events where at one time two particles are
created, and at the other time two are destroyed. All the other
terms in the correlation function of the four $B$'s will vanish.
$s_o$ and $s_t$ are given by
\begin{eqnarray}
s_o = {1\over 16} &\bigg[&c_{\omega_1} c_{\omega_2} c_{\omega_3}
c_{\omega_4}  \langle
     A_{-\omega_1}^\dagger A_{\omega_2} A_{-\omega_3}^\dagger
A_{\omega_4}  \rangle   \nonumber\\
  &+& c_{\omega_1} c_{\omega_2} d_{\omega_3} d_{\omega_4}
 \langle
     A_{-\omega_1}^\dagger A_{\omega_2} A_{\omega_3} A^\dagger_{-
\omega_4}  \rangle   \nonumber\\
  &+& d_{\omega_1} d_{\omega_2} c_{\omega_3} c_{\omega_4}
 \langle
     A_{\omega_1} A^\dagger_{-\omega_2} A^\dagger_{-\omega_3}
A_{\omega_4}  \rangle   \nonumber\\
  &+& d_{\omega_1} d_{\omega_2} d_{\omega_3} d_{\omega_4}
 \langle
     A_{\omega_1} A^\dagger_{-\omega_2} A_{\omega_3} A^\dagger_{-
\omega_4}  \rangle \bigg],
\label{Sodef}
\end{eqnarray}
and
\begin{eqnarray}
s_t = {1\over 16} &\bigg[&d_{\omega_1} c_{\omega_2} c_{\omega_3}
d_{\omega_4}  \langle
     A_{\omega_1} A_{\omega_2} A_{-\omega_3}^\dagger A^\dagger_{-
\omega_4}  \rangle  \nonumber\\
  &+& c_{\omega_1} d_{\omega_2} d_{\omega_3} c_{\omega_4}
 \langle
     A_{-\omega_1}^\dagger A^\dagger_{-\omega_2} A_{\omega_3}
A_{\omega_4}  \rangle \bigg].
\label{Stdef}
\end{eqnarray}
In these equations, $c_\omega$ and $d_\omega$ are given by
\begin{equation}
c_\omega = 1 + e^{i\phi(\omega)} \qquad {\rm and} \qquad
d_\omega = 1 - e^{i\phi(\omega)},
\label{cddef}
\end{equation}
with $\phi(\omega)$ defined in Eq. (\ref{phidef}). The correlations
of the four $A$'s can be evaluated using Eqs. (\ref{Acom}) and
(\ref{AdAexp}).  If we interchange $\omega_1$ with $\omega_2$ in
the second two lines of $s_o$, and perform the sums over $\omega_3$
and $\omega_4$, we obtain
\begin{eqnarray}
\sum_{\omega_3, \omega_4}
           s_o\ e^{i(\omega_1+\omega_2+\omega_3+\omega_4)x_+}
= {1\over 16}&\bigg[&c_{\omega_1} c_{\omega_2} c_{-\omega_1} c_{-
\omega_2}                   - c_{\omega_1} c_{\omega_2} d_{-
\omega_1} d_{-\omega_2}           \nonumber\\
                 &-& d_{\omega_1} d_{\omega_2} c_{-\omega_1} c_{-
\omega_2}                   + d_{\omega_1} d_{\omega_2} d_{-
\omega_1} d_{-\omega_2}                    \bigg]  \nonumber\\
                  &\times& n_{-\omega_1}(1-n_{\omega_2}).
\label{Socalc}
\end{eqnarray}
In this equation, the expression containing the number operators is
the same as for the noise in the incoming current, so we will
obtain the same  limits of integration as in equation
(\ref{nnlimits}).
Next, we can expand out the $c_\omega$'s and $d_\omega$'s in terms
of $\omega$ and substitute this back into equation (\ref{Sjjdef})
for the noise. After performing the integrals over $t$ and
$\omega_1$, we find that the  contribution to the noise due to
$s_o$ has the form
\begin{equation}
S_o(\omega;x_+,x_+) = \int_{\omega_0}^{\omega_0 \pm \omega}
\frac{d\omega_2}{2\pi}          \ {\left((4\pi|\Gamma|^2)^2 -
\omega_2(\omega_2 \mp \omega_1)\right)^2            \over
\left((\omega_2 \mp \omega)^2 + (4\pi|\Gamma|^2)^2\right)
         \left(\omega_2^2 + (4\pi|\Gamma|^2)^2\right)}
\theta(\pm \omega),
\label{Soppint}
\end{equation}
where we sum over the two different signs in front of $\omega$.
Upon performing the $\omega_2$ integral, we obtain
\begin{eqnarray}
S_o(\omega;x_+,x_+) =  \frac{|\omega|}{2\pi} &-& 2 |\Gamma|^2
  \left[\tan^{-1}\left({|\omega|-\omega_0 \over4\pi |\Gamma|^2}\right)
  + \tan^{-1}\left({|\omega|+\omega_0 \over4\pi |\Gamma|^2}\right)\right]
\nonumber\\
    &-& 8\pi{|\Gamma|^4 \over |\omega|}\bigg[
            2\ln\left(\omega_0^2 + (4\pi |\Gamma|^2)^2 \right)
     -   \ln\left((\omega + \omega_0)^2 + (4\pi |\Gamma|^2)^2 \right)
\nonumber\\
&\ & \ \ \ \ \ \ \ \ \ \ -   \ln\left((\omega - \omega_0)^2 + (4\pi
|\Gamma|^2)^2 \right)\bigg].
\label{Sopp}
\end{eqnarray}

Next, we will calculate the contribution to the noise due to $s_t$.
The two expectation values we must evaluate are $\langle
A_{\omega_1} A_{\omega_2} A_{-\omega_3}^\dagger A^\dagger_{-
\omega_4}\rangle$ and $\langle
A_{-\omega_1}^\dagger A^\dagger_{-\omega_2} A_{\omega_3}
A_{\omega_4} \rangle$. In both cases, either $\omega_1$ is paired
with $\omega_3$ and $\omega_2$ is paired with $\omega_4$, or
$\omega_1$ is paired with $\omega_4$ and $\omega_2$ with
$\omega_3$.  Thus we have
\begin{equation}
 \langle
     A_{\omega_1} A_{\omega_2} A_{-\omega_3}^\dagger A^\dagger_{-
\omega_4}  \rangle
  = (1- n_{\omega_1})(1-n_{\omega_2})
    (\delta_{\omega_1,-\omega_4}\delta_{\omega_2,-\omega_3}
- \delta_{\omega_1,-\omega_3}\delta_{\omega_2,-\omega_4}),
\label{AAAdAd}
\end{equation}
and
\begin{equation}
 \langle
     A_{-\omega_1}^\dagger A^\dagger_{-\omega_2} A_{\omega_3}
A_{\omega_4}  \rangle
  =  n_{-\omega_1} n_{-\omega_2}
    (\delta_{-\omega_1,\omega_4}\delta_{-\omega_2,\omega_3}
- \delta_{-\omega_1,\omega_3}\delta_{-\omega_2,\omega_4}).
\label{AdAdAA}
\end{equation}
Substituting these expressions into the equation for $s_t$ and
performing the integrals over $\omega_3$ and $\omega_4$, we obtain
\begin{eqnarray}
\sum_{\omega_3, \omega_4} s_t\
e^{i(\omega_1+\omega_2+\omega_3+\omega_4)x_+}   = {1\over16} &\
&\left[d_{\omega_1} d_{-\omega_1} c_{\omega_2} c_{-\omega_2}
               -d_{\omega_1} c_{-\omega_1} c_{\omega_2} d_{-
\omega_2}                   \right] \nonumber\\
  &\times&\left[(1-n_{\omega_1})(1-n_{\omega_2}) + n_{-
\omega_1}n_{-\omega_2}                   \right].
\label{Stcalc}
\end{eqnarray}
When we expand the $c$'s and $d$'s in terms of $\omega$ and perform
the integral over $t$, we find that the contribution to the noise
due to $s_t$ is given by
\begin{eqnarray}
S_t(\omega;x_+,x_+)
    = &\int& \frac{d\omega_1}{2\pi} \frac{d\omega_2}{2\pi}
    {(4\pi|\Gamma|^2)^2(\omega_2^2 - \omega_1 \omega_2) \over
    (\omega_1^2 + (4\pi|\Gamma|^2)^2)(\omega_2^2 + (4\pi|\Gamma|^2)^2)}
\nonumber\\
 &\ & \ \ \times \delta(\omega_1 + \omega_2 \pm\omega)
  \left[(1-n_{\omega_1})(1-n_{\omega_2}) + n_{-\omega_1}n_{- \omega_2}
    \right] ,
\label{Stppcalc}
\end{eqnarray}
where again it is understood that we sum the two integrands with
the different sign in front of $\omega$.  After the integration
over $\omega_1$ is performed, the expression in square brackets
becomes
\begin{equation}
(1-n_{-\omega_2\mp\omega})(1-n_{\omega_2})
    + n_{\omega_2\pm\omega}n_{-\omega_2}
    = \cases{1 & for $\omega_0 < \omega_2 < \mp\omega-\omega_0$
                  and $\mp\omega-2\omega_0 > 0$ \cr
             1 & for $-\omega_0 < \omega_2 < \mp\omega+\omega_0$
                   and $\mp\omega+2\omega_0 > 0$ \cr
             0 & otherwise. }
\label{nlimitstwo}
\end{equation}
We note that this time the limits of integration determined by the
factors of $n$ impose cutoffs at $\omega = \pm 2\omega_0$.  These
are the origins of the singularities at $\omega = 2\omega_0$,
which, as we shall
see shortly, persist for all $|\Gamma| \ne 0$.  After equation
(\ref{nlimitstwo}) is substituted into the equation for
$S_t(\omega;x_+,x_+)$, the noise becomes
\begin{equation}
S_t(\omega;x_+,x_+)= \sum_{a,b = \pm1} \theta(a\omega + b2\omega_0)
\int_{-b\omega_0}^{b\omega_0 - a\omega} \frac{d\omega_2}{2\pi}
{(4\pi|\Gamma|^2)^2\left(\omega_2^2 +  \omega_2(\omega_2 +
a\omega)\right) \over          \left((\omega_2 + a\omega)^2 +
(4\pi|\Gamma|^2)^2\right)          \left(\omega_2^2 +
(4\pi|\Gamma|^2)^2\right)}.
\label{Stppint}
\end{equation}
The integration over $\omega_2$ yields
\begin{eqnarray}
S_t(\omega;x_+,x_+) = \sum_{a,b = \pm1} \theta&(&a\omega + b2\omega_0)
   \bigg\{2|\Gamma|^2\left[\tan^{-1}\left({b\omega_0\over 4\pi
|\Gamma|^2}\right)
             +\tan^{-1}\left({a\omega
+b\omega_0\over 4\pi |\Gamma|^2}\right)
          \right]
\nonumber\\
     &+&{8\pi|\Gamma|^4 \over a\omega}\left[\ln\left((4\pi
|\Gamma|^2)^2 +\omega_0^2\right)
             -\ln\left((4\pi
|\Gamma|^2)^2 +(a\omega + b\omega_0)^2\right)\right]
       \bigg\}.
\label{Stpp}
\end{eqnarray}
We note that this contribution to the noise has the step function
which provides a ``sharp'' singularity at $|\omega| = |2\omega_0|$,
for any non-zero value of $|\Gamma|$.  This is the electron
singularity.  However, for $|\Gamma| \to 0$, the arctangents
provide a singularity at $|\omega| = |\omega_0|$, which is the
quasiparticle singularity.

When we add the two contributions to the noise together, we find
that the noise on the outgoing side of the impurity is
\begin{eqnarray}
S(\omega;x_+,x_+) = \frac{|\omega|}{2\pi} + &\theta&(2|\omega_0| -
|\omega|)     \bigg\{4
|\Gamma|^2\left[\tan^{-1}\left({|\omega_0|\over 4\pi
|\Gamma|^2}\right)             +\tan^{-1}\left({|\omega_0| -
|\omega|\over 4\pi |\Gamma|^2}\right)          \right]  \nonumber\\
    &+&16\pi{|\Gamma|^4\over |\omega|}
     \left[\ln\left((4\pi |\Gamma|^2)^2 +(|\omega| -
|\omega_0|)^2\right)             -\ln\left((4\pi |\Gamma|^2)^2
+\omega_0^2\right)\right]       \bigg\}.
\label{Sppc}
\end{eqnarray}

Finally, the noise between the currents on either side of the
impurity can be calculated similarly, so we will omit the details
here.


\newpage

\begin{center}
{\large FIGURE CAPTIONS}
\end{center}

\

Figure 1. Geometries for tunneling between edge states. By adjusting
the gate voltage $V_G$ one can obtain either a simply connected QH
droplet (a), or two disconnected QH droplets (b). For the geometry in
(a) both electrons and quasiparticles (carrying fractional charge) can
tunnel from one edge to the other, whereas for the tunneling geometry
in (b) only electrons can tunnel. The Luttinger liquid behavior is
characterized by the exponent $g=\nu$ in (a), and $g=\nu^{-1}$ in (b).
The tunneling current $I_t$ depends on the applied voltage between the
right and left edges, and by increasing this voltage one can also
cross over from the geometry (b) to the geometry (a).

\

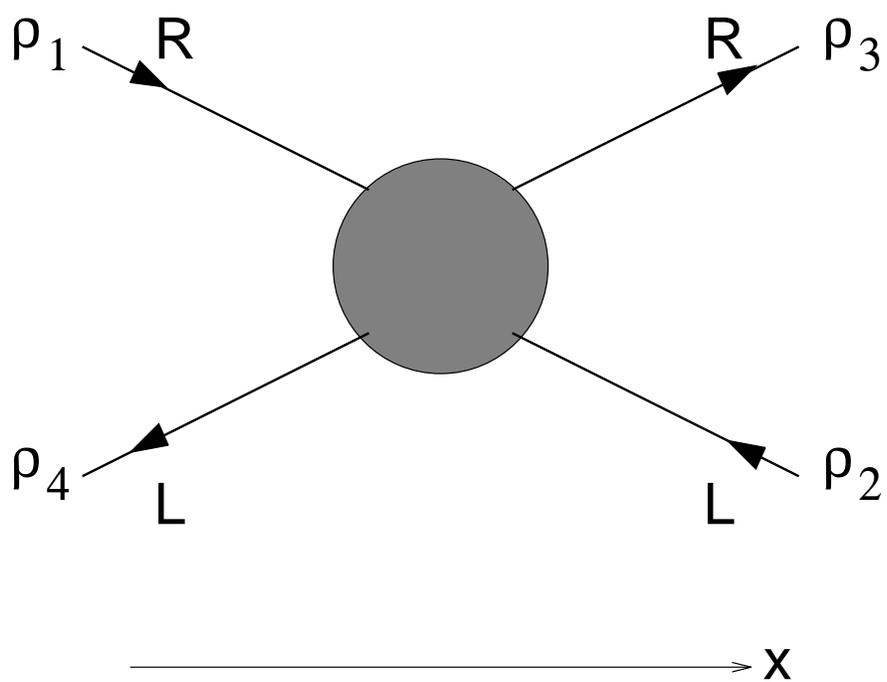
Figure 2. Four terminal geometry for the measurement of tunneling
between edge states. The terminals 1 and 2 correspond to branches
that are incoming to the scatterer, while terminals 3 and 4
correspond to outgoing ones. The arrows indicate the direction of
propagation for a given branch. The incoming branches are in
equilibrium with their reservoirs of origin, while the outgoing
ones do get effected by the scatterer. By measuring fluctuations in
the voltages/currents at the four terminals ($V_i/I_i$,
$i=1,2,3,4$), the auto-correlation spectra $S_{ij}(\omega)$, with
$i=j$, and the cross-correlation spectra $S_{ij}(\omega)$, with
$i\ne j$, can be obtained. These voltage/current fluctuations
contain information on the fluctuations of the tunneling current.

\

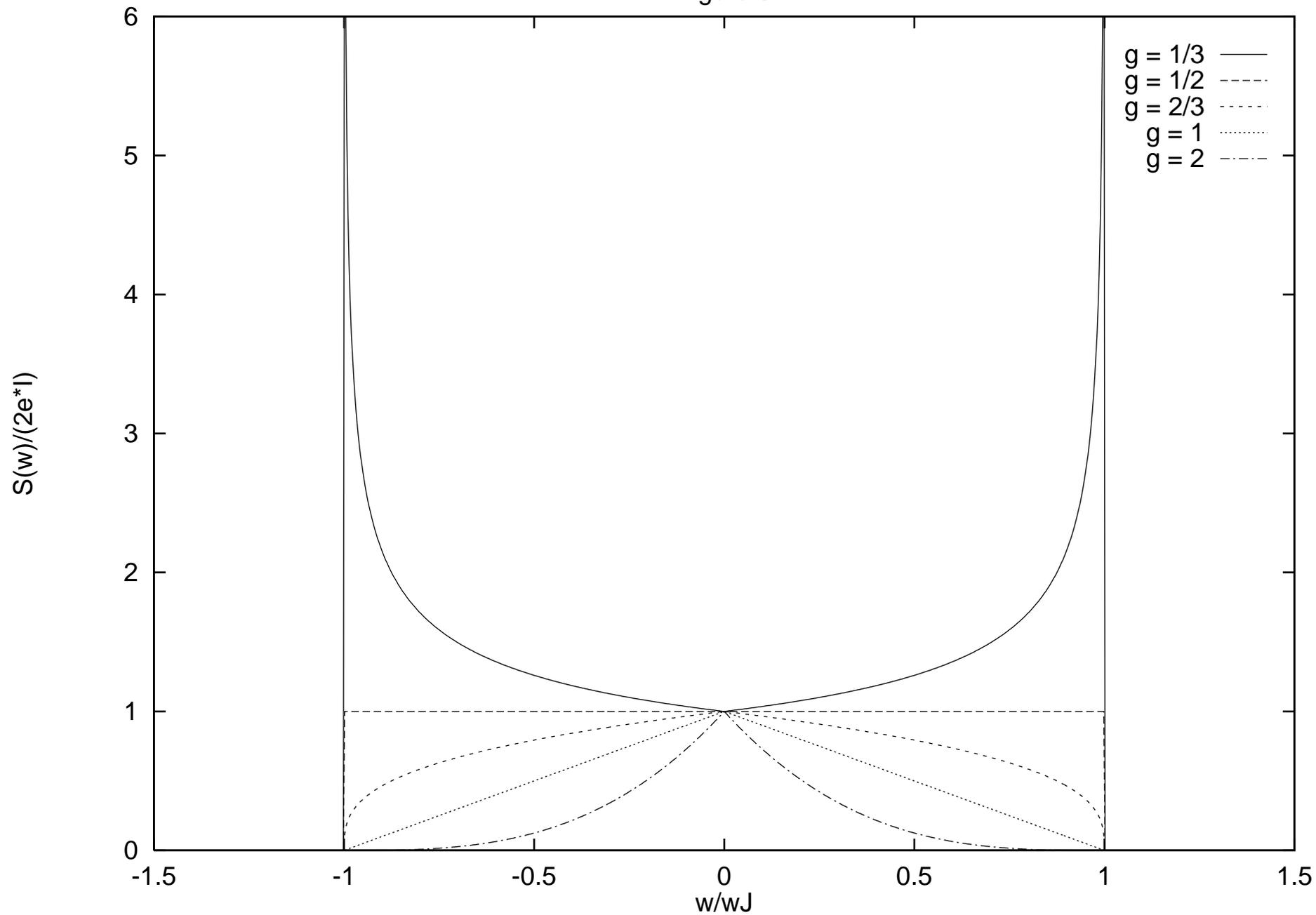
Figure 3. Plots of the excess noise of outgoing branches (probes 3
and 4 of Fig. 2) calculated to second order in perturbation theory
(equation (\ref{noise2ndorder})). The excess noise is normalized to
the zero-
frequency shot noise level, and the frequency $\omega$ to the
Josephson frequency $\omega_J$ ($S^{(2)}(\omega)/2e^*I_t$ {\it vs.}
$\omega/\omega_J$). Different singularities are obtained at
$\omega=\omega_J$ for different values of $g$:
${1\over3}$,  ${1\over2}$, ${2\over3}$,  $1$, and  $2$. One should
keep in mind that, although the singularities all occur at
$\omega=\omega_J$, the value of $\omega_J$ depends on the charge of
the current carrier, which in turn also depends on $g$.

\

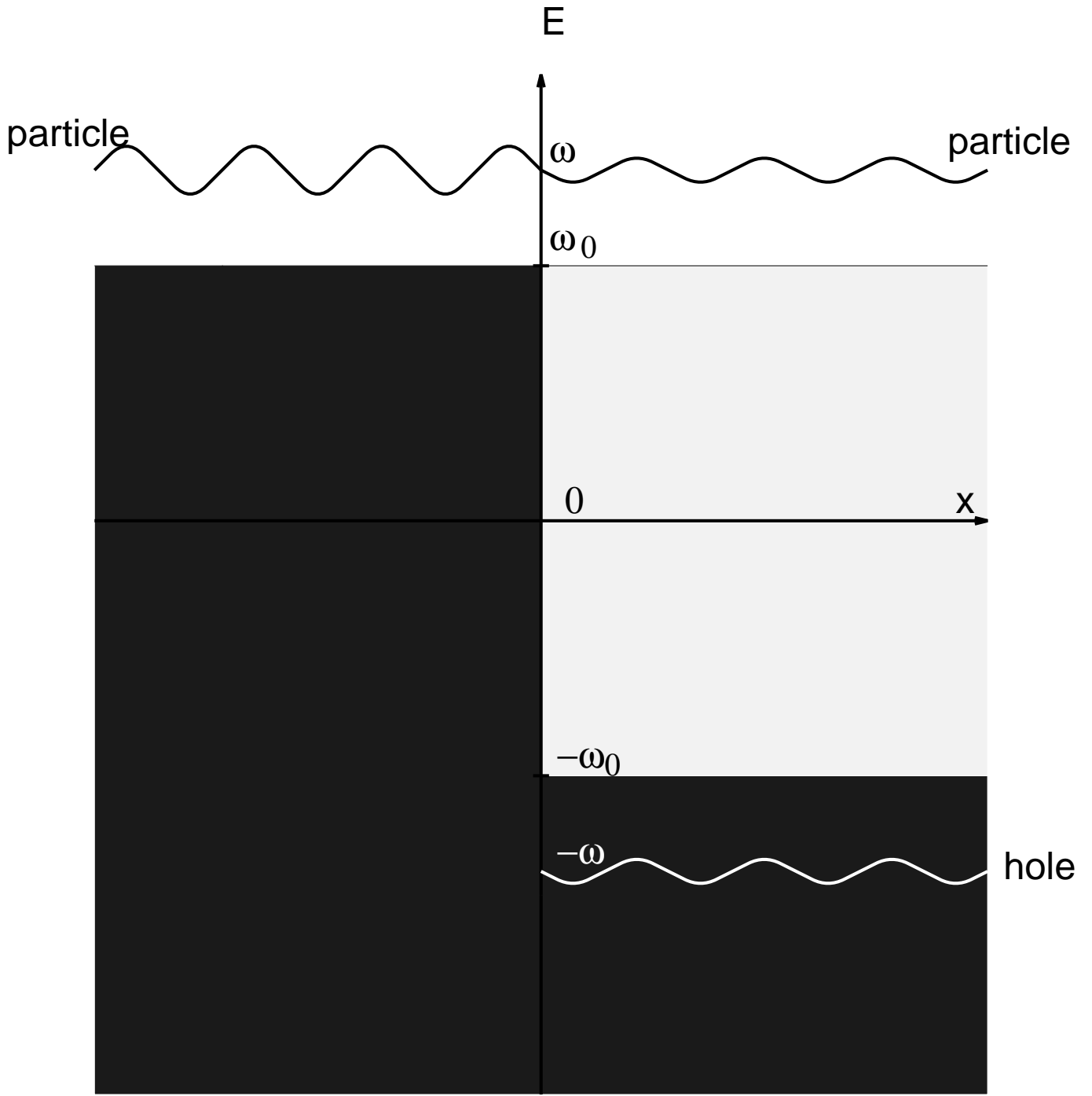
Figure 4. A particle (plane wave) incoming from the left ($x<0$) with
energy $\omega$ scatters off the impurity at $x=0$ into a
superposition of a particle at energy $\omega$ and a hole at energy
$-\omega$ on the right side of the impurity ($x>0$). In the case where
the incoming state is a filled Fermi sea up to the energy $\omega_0$,
the scattered state on the right side of the impurity will be
completely filled up to energy $-\omega_0$, and partially filled
between $-\omega_0$ and $\omega_0$. It is this partially filled energy
range from $-\omega_0$ and $\omega_0$ which is responsible for the
non-equilibrium properties of the system.

\

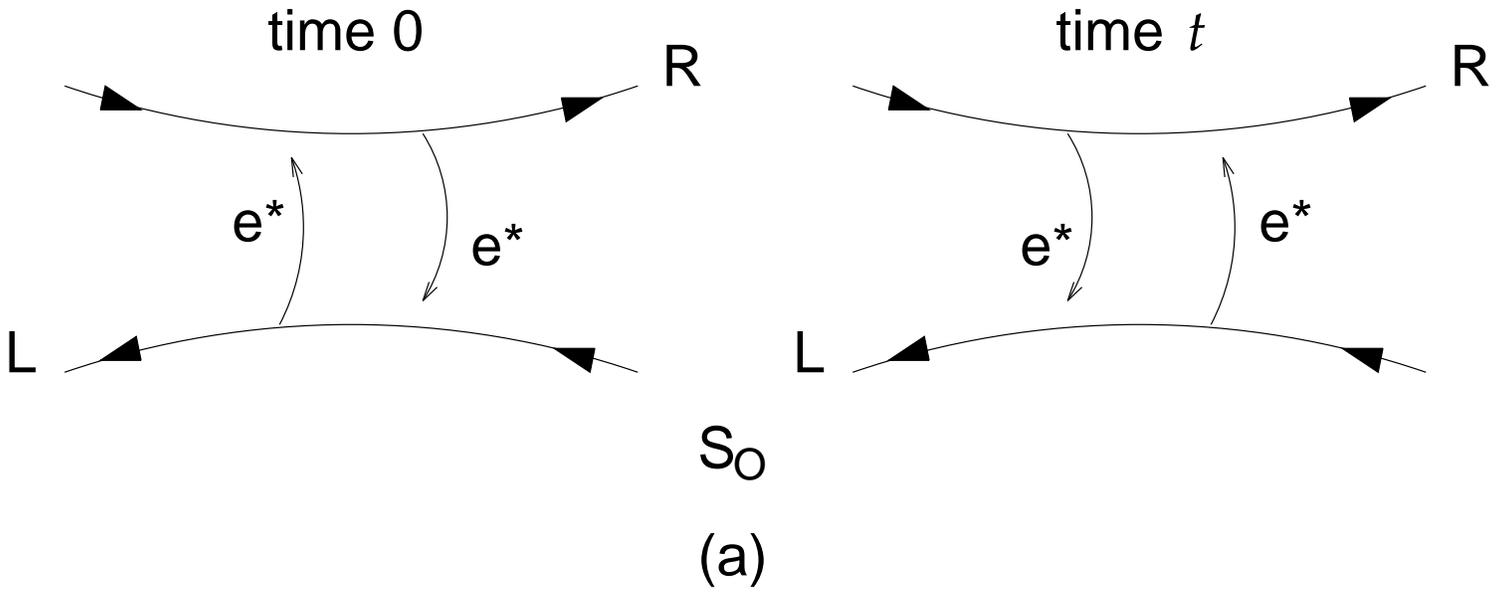
Figure 5. The tunneling processes $s_0$ (a) and $s_t$ (b). In the
process $s_0$, both at time 0 and $t$, a quasiparticle tunnels from
the left branch to the right branch, and another quasiparticle tunnels
in the opposite direction. In the $s_t$ process, at time 0 two
quasiparticles tunnel from, say, the left to the right branch, and at
time $t$ the two quasiparticles tunnel back in the opposite direction.
The process $s_t$ is responsible for the singularity at the electron
frequency $\tilde{\omega}_0=2\omega_0$.

\

Figure 6. Plots for the renormalized noise
$\tilde{S}$ {\it vs.}
$\tilde\omega/\tilde\omega_0$. $\tilde{S}$, $\tilde{\omega}$ and
$\tilde{\omega}_0$ are the renormalized noise and frequencies, using
the coupling constant as the scaling factor ($\tilde S = {S \over
2|\Gamma|^2}$,
$\tilde \omega = {\omega\over 4\pi|\Gamma|^2}$ and $\tilde \omega_0 =
{\omega_0\over 4\pi|\Gamma|^2}$).  In (a) the excess noise
$\tilde S - \tilde S^{\tilde \omega_0 = 0}$ is plotted for large
values of $\tilde \omega_0$, which illustrate the weak coupling
($|\Gamma| \to 0$) limit.  The rescaled excess noise
$(\tilde S - \tilde S^{\tilde
\omega_0 = 0})/\tilde \omega_0^3$ is plotted in (b).  It shows the
strong coupling limit ($|\Gamma| \to \infty$) as $\omega_0 \to 0$.
The full noise $\tilde S/\tilde\omega_0$ is plotted in figure (c).
For the larger values of $\omega_0$, notice that the singularity
at $\tilde \omega = 2\tilde\omega_0$ is hidden in the full noise.
Meanwhile, some reminiscent signs of the quasiparticle singularity appear
near $\tilde \omega = \tilde \omega_0$.

\

Figure 7. The association of the four densities $\rho_i$ ($i=1,2,3,4$)
to the left and right moving branches for the dual pictures
corresponding to (a) $g=\nu$ and (b) $g=\nu^{-1}$ (compare to Figs. 1a
and 1b). Notice that $\rho_3$ and $\rho_4$ change chirality under
duality, and that the space coordinates (the $x$ and $\tilde x$ axis)
should also be redefined under the duality transformation.

\

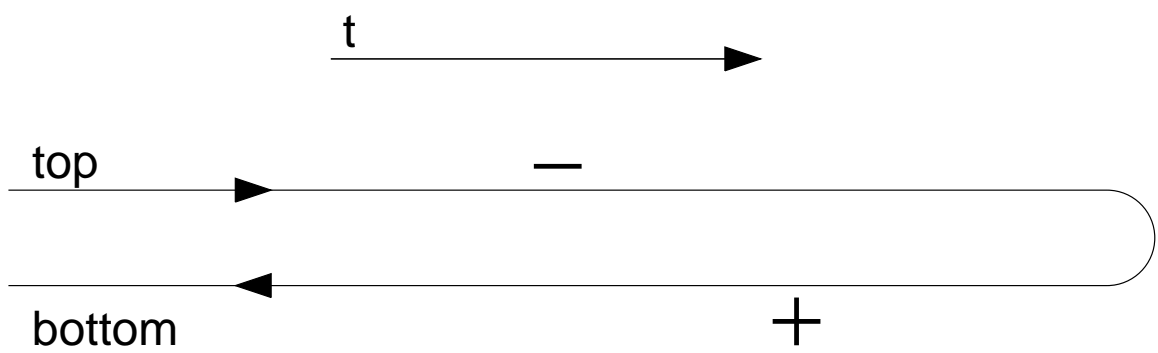
Figure 8. An insertion of an operator $e^{+iq\phi(t)}$ corresponds
to the insertion of a charge $+$ on the contour at time $t$.
Similarly, an insertion of an operator $e^{-iq\phi(t)}$ corresponds
to an insertion of a charge $-$ at time $t$. The time $t$ is
ordered along the contour shown, and there is a distinction between
charges placed on the top and bottom branches. In the illustration,
we consider the particular case when the $-$ charge is inserted on
the top contour, and the $+$ charge is inserted on the bottom
contour.

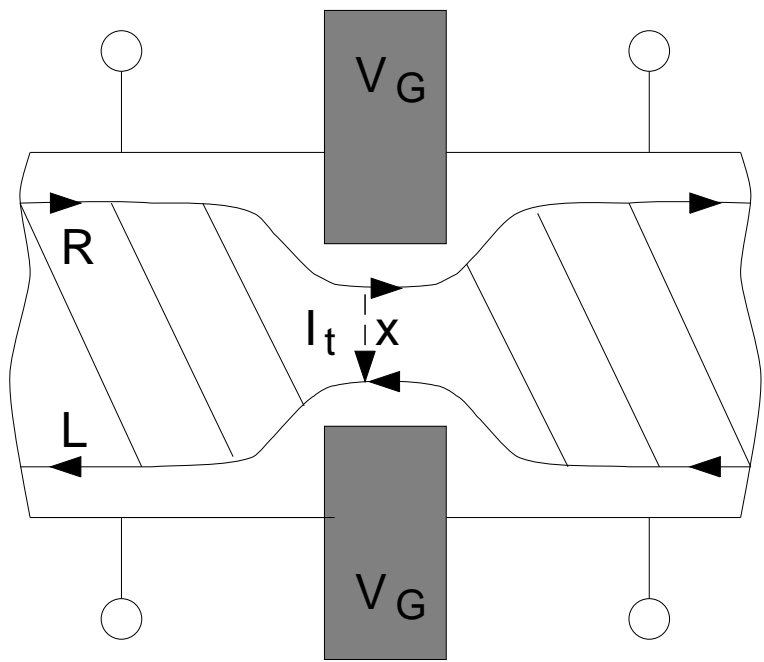
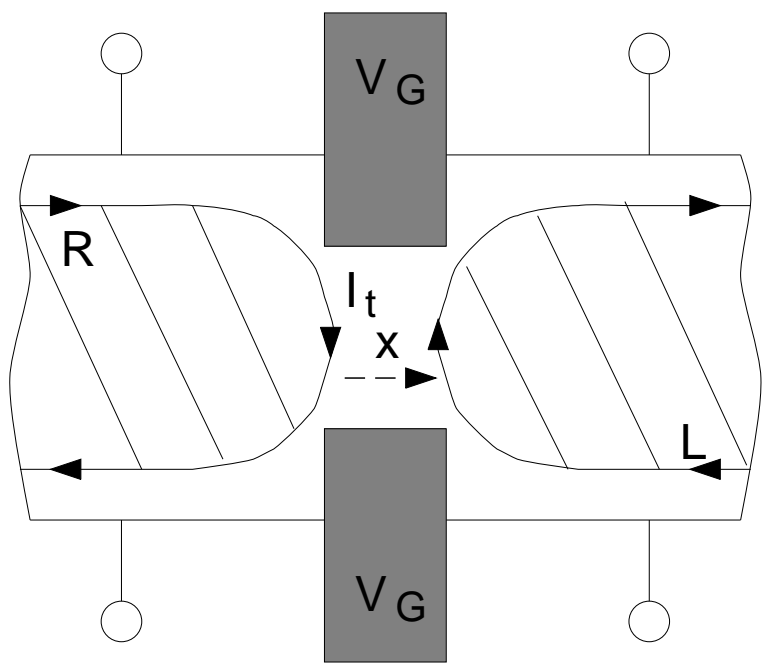
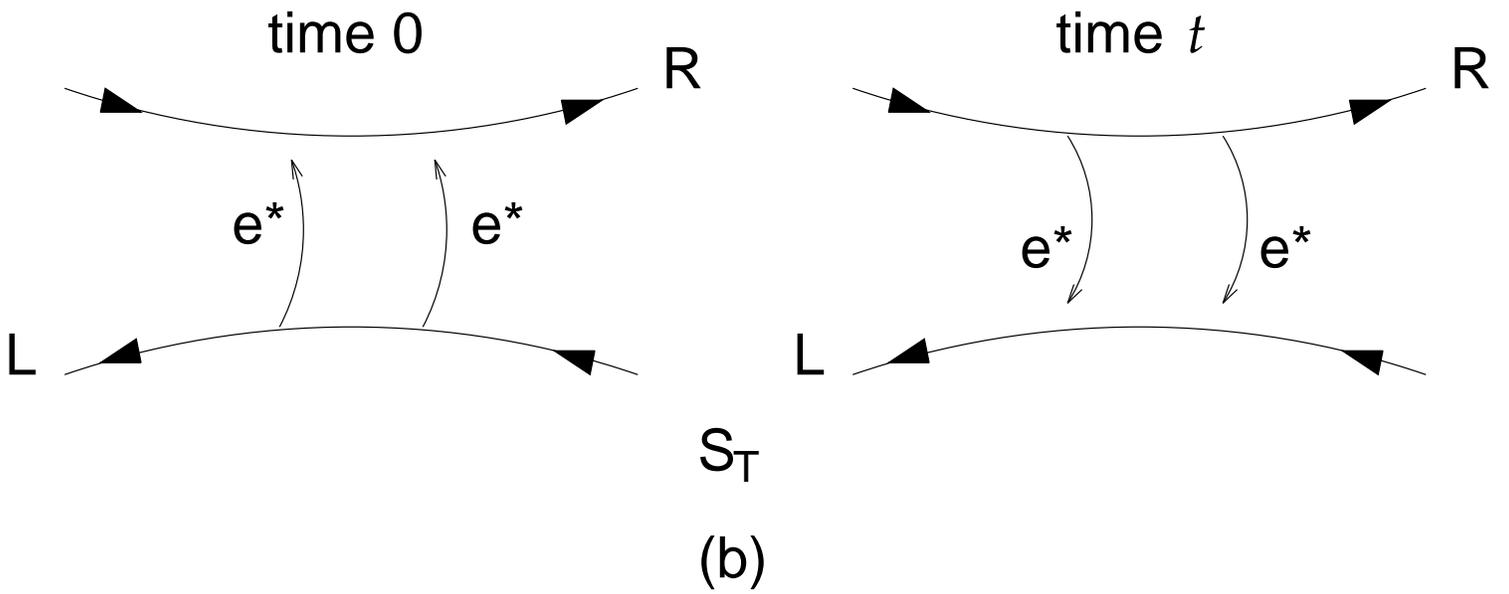
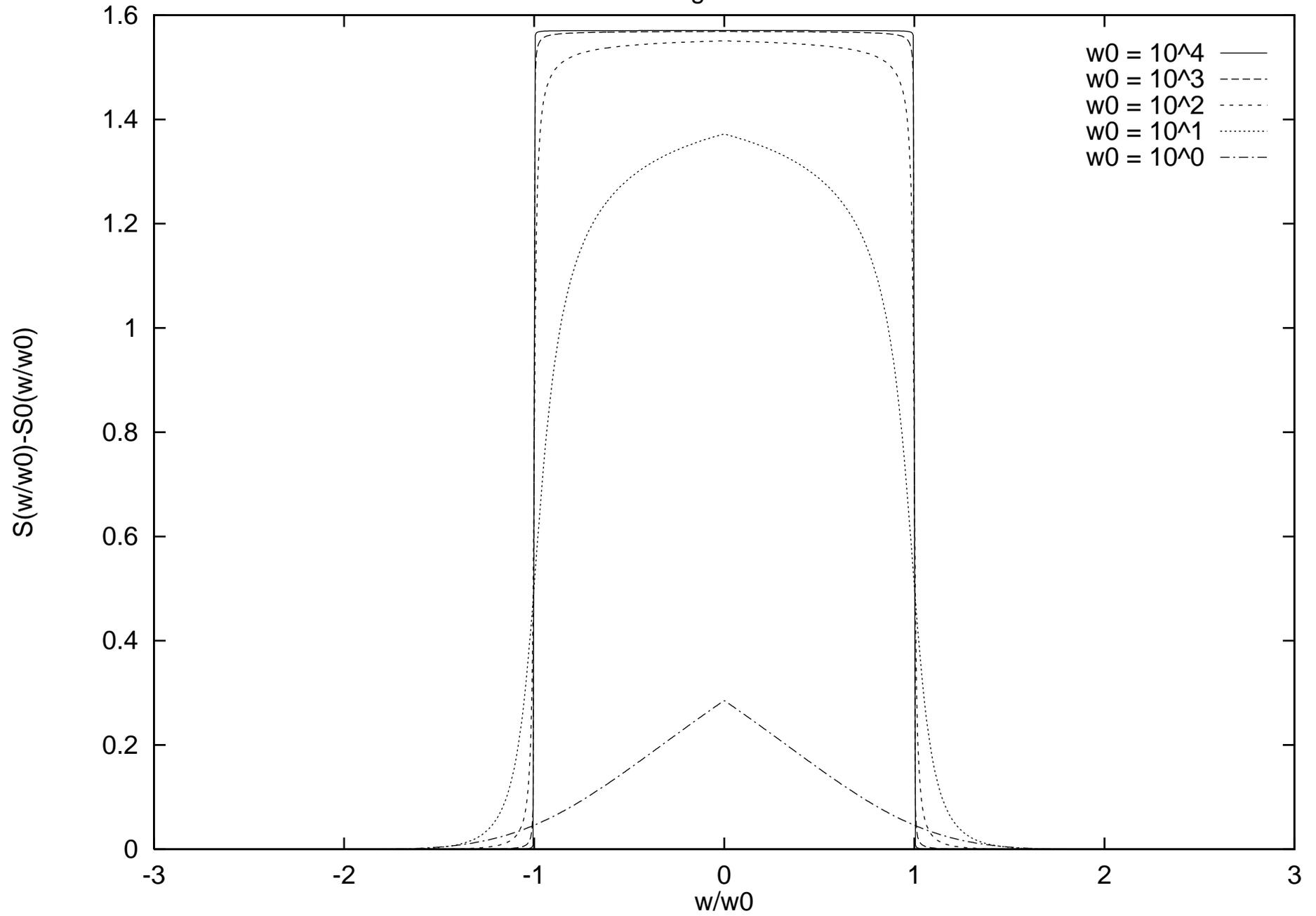
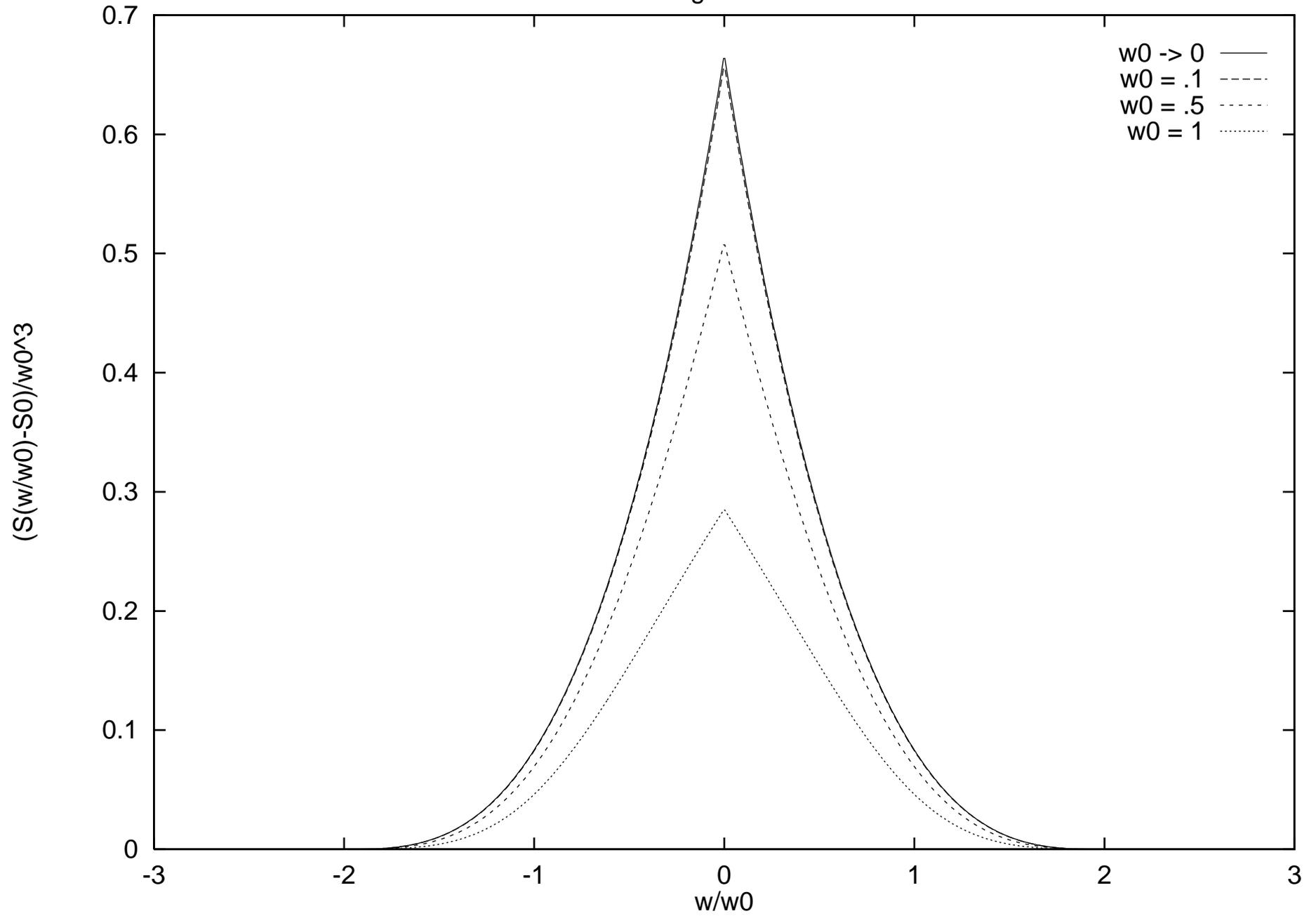
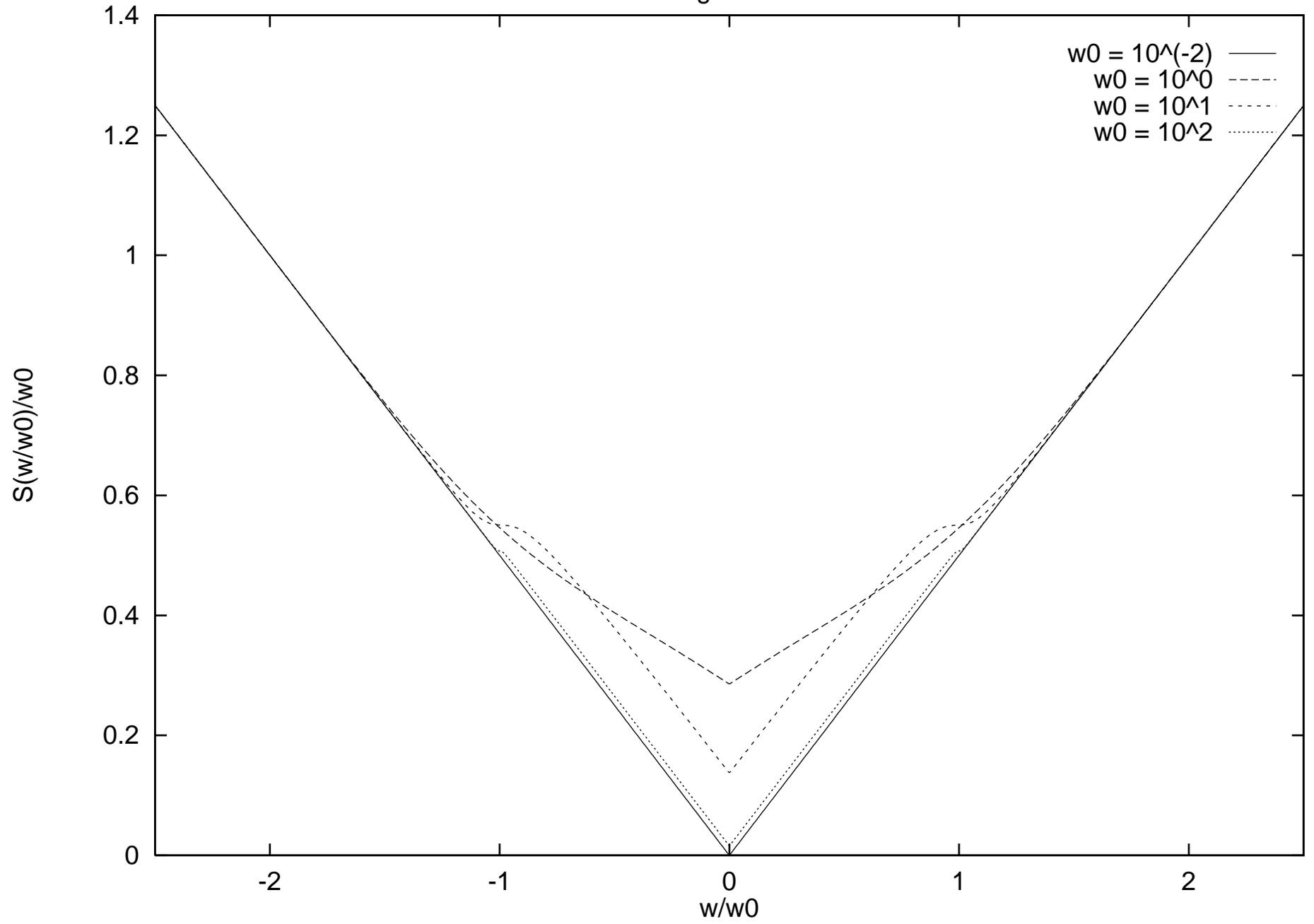
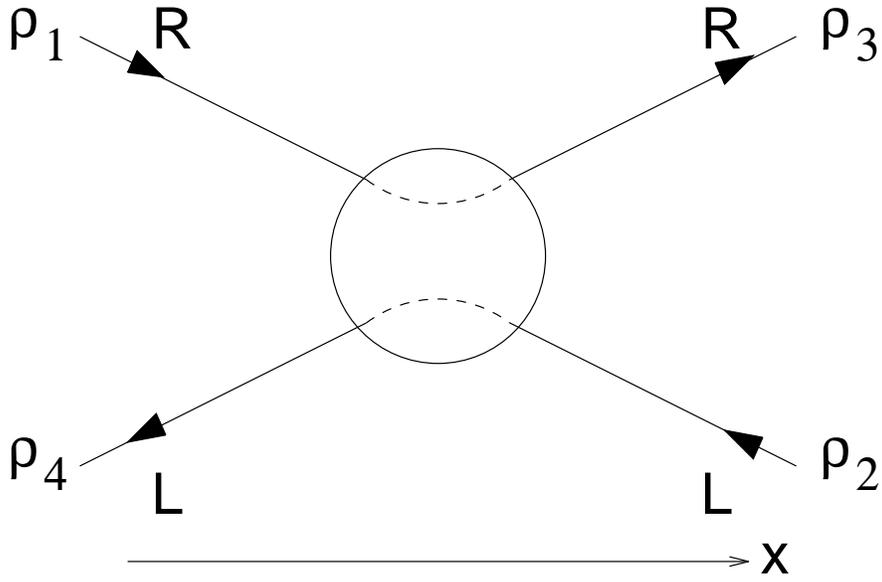
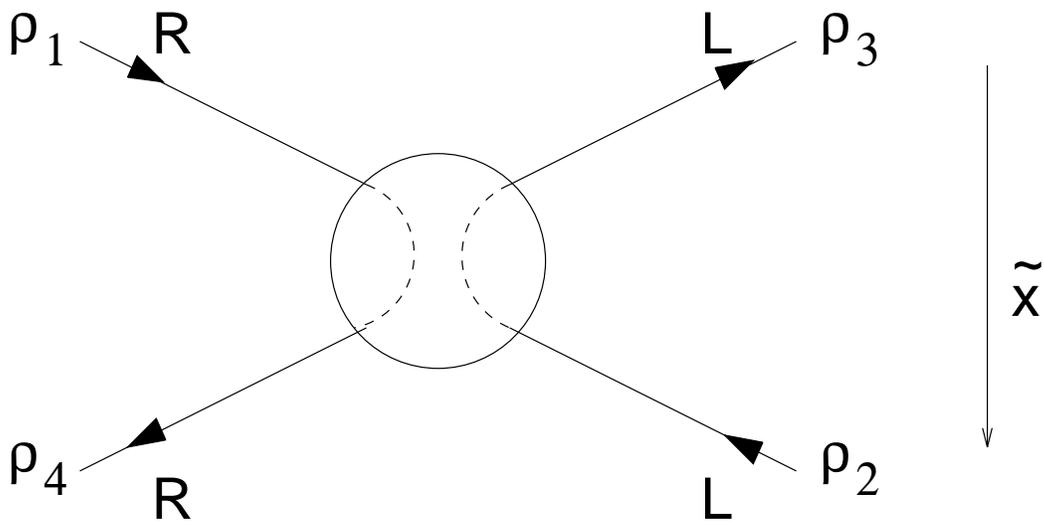
\end{document}